\documentclass[3p,times]{elsarticle}

\usepackage{caption}
\usepackage{multicol}
\usepackage{graphicx}
\usepackage{subfig}
\usepackage{tabularx}
\usepackage{latexsym}
\usepackage{amstext,amsfonts,amssymb,amsmath}
\usepackage{todonotes}
\usepackage[version=3]{mhchem}
\usepackage{epstopdf}
\AppendGraphicsExtensions{.tif}
\usepackage{multicol}
\usepackage{nonfloat}
\usepackage{color}
\usepackage{bm}
\usepackage{algorithm}
\usepackage{algpseudocode}
\usepackage{multirow}
\usepackage{tikz}
\usepackage{amsmath}
\usepackage{amssymb}
\usepackage{breqn}
\usepackage{etoolbox}

\newrobustcmd*{\myVtriangle}[2]{\tikz{\filldraw[draw=#1,fill=#2] (0cm,0.2cm) --
(0.2cm,0.2cm) -- (0.1cm,0cm) -- (0cm,0.2cm);}}

\newrobustcmd*{\mythickVtriangle}[2]{\tikz{\filldraw[line width=0.3mm,draw=#1,fill=#2] (0cm,0.2cm) --
(0.2cm,0.2cm) -- (0.1cm,0cm) -- (0cm,0.2cm);}}

\newrobustcmd*{\mythickErrorVtriangle}[2]{\tikz{\filldraw[line width=0.3mm,draw=#1,fill=#2] (-0.05cm,0.05cm) --
(0.05cm,0.05cm) -- (0cm,-0.05cm) -- (-0.05cm,0.05cm);  \draw[draw=#1] (0.0cm, -0.12cm) -- (0.0cm, 0.12cm) ; \draw[draw=#1] (-0.06cm, 0.12cm) -- (0.06cm, 0.12cm); \draw[draw=#1] (-0.06cm, -0.12cm) -- (0.06cm, -0.12cm)    }}

\newrobustcmd*{\mytriangle}[2]{\tikz{\filldraw[draw=#1,fill=#2] (0.0cm,0.0cm) --
(0.2cm,0cm) -- (0.1cm,0.2cm) -- (0cm,0cm);}}

\newrobustcmd*{\mysquare}[2]{\tikz{\draw[draw=#1,fill=#2] (0cm,0cm)
rectangle (0.2cm,0.2cm)}}

\newrobustcmd*{\mythicktriangle}[2]{\tikz{\filldraw[line width=0.3mm,draw=#1,fill=#2] (0.0cm,0cm) --
(0.2cm,0cm) -- (0.1cm,0.2cm) -- (0.0cm,0cm);}}

\newrobustcmd*{\mythicksquare}[2]{\tikz{\draw[line width=0.3mm,draw=#1,fill=#2] (0cm,0cm)
rectangle (0.2cm,0.2cm)}}

\newrobustcmd*{\mybarredtriangle}[2]{\tikz{\draw[draw=#1,fill=#2] (0,0) --
(0.2cm,0) -- (0.1cm,0.2cm) -- (0cm,0cm); \draw[draw=#1] (-0.1cm, 0.07cm) -- (0.3cm, 0.07cm)}}

\newrobustcmd*{\mythickbarredtriangle}[2]{\tikz{\draw[line width=0.3mm,draw=#1,fill=#2] (0,0) --
(0.2cm,0) -- (0.1cm,0.2cm) -- (0cm,0cm); \draw[draw=#1] (-0.1cm, 0.07cm) -- (0.3cm, 0.07cm)}}

\newrobustcmd*{\mybarredsquare}[2]{\tikz{\draw[draw=#1,fill=#2] (0,0)
rectangle (0.2cm,0.2cm); \draw[draw=#1] (-0.1cm, 0.1cm) -- (0.3cm, 0.1cm)}}

\newrobustcmd*{\mythickbarredsquare}[2]{\tikz{\draw[line width=0.3mm,draw=#1,fill=#2] (0,0)
rectangle (0.2cm,0.2cm); \draw[draw=#1] (-0.1cm, 0.1cm) -- (0.3cm, 0.1cm)}}

\newrobustcmd*{\mybarredcircle}[2]{\tikz{\draw[draw=#1,fill=#2] (0,0)
circle (0.1cm); \draw[draw=#1] (-0.2cm, 0.0cm) -- (0.2cm, 0.0cm)}}

\newrobustcmd*{\mythickbarredcircle}[2]{\tikz{\draw[line width=0.3mm,draw=#1,fill=#2] (0,0)
circle (0.1cm); \draw[draw=#1] (-0.2cm, 0.0cm) -- (0.2cm, 0.0cm)}}

\newrobustcmd*{\mythickErrorcircle}[2]{\tikz{\draw[line width=0.3mm,draw=#1,fill=#2] (0,0)
circle (0.06cm); \draw[draw=#1] (0.0cm, -0.12cm) -- (0.0cm, 0.12cm) ;   \draw[draw=#1] (-0.06cm, 0.12cm) -- (0.06cm, 0.12cm); \draw[draw=#1] (-0.06cm, -0.12cm) -- (0.06cm, -0.12cm)    }}

\newrobustcmd*{\mydashedline}[1]{\tikz{\draw[draw=#1] (-0.2cm, 0.2cm) -- (-0.1cm, 0.2cm); \draw[draw=#1] (-0.0cm, 0.2cm) -- (0.1cm, 0.2cm)}}

\newrobustcmd*{\mythickcross}[1]{\tikz{\draw[line width=0.3mm,draw=#1] (0,0) --
(0.2cm,0); \draw[line width=0.3mm,draw=#1] (0.1cm,-0.1cm) -- (0.1cm,0.1cm);}}

\newrobustcmd*{\mybarredcross}[1]{\tikz{\draw[line width=0.3mm,draw=#1] (0,0) --
(0.2cm,0); \draw[line width=0.3mm,draw=#1] (0.1cm,-0.1cm) -- (0.1cm,0.1cm); \draw[draw=#1] (-0.1cm,0) -- (0.3cm,0);}}

\newrobustcmd*{\myline}[1]{\tikz{\draw[draw=#1] (-0.15cm, 0.1cm) -- (0.15cm, 0.1cm);\draw[line width=0.3mm,draw=#1] (-0.0cm, 0.0cm);}}

\newrobustcmd*{\mythickline}[1]{\tikz{\draw[line width=0.3mm,draw=#1] (-0.15cm, 0.1cm) -- (0.15cm, 0.1cm);\draw[line width=0.3mm,draw=#1] (-0.0cm, 0.0cm);}}

\newrobustcmd*{\mythickdashedline}[1]{\tikz{\draw[line width=0.3mm,draw=#1] (-0.2, 0.1cm) -- (-0.1cm, 0.1cm); \draw[line width=0.3mm,draw=#1] (-0.0cm, 0.1cm) -- (0.1cm, 0.1cm); \draw[line width=0.3mm,draw=#1] (-0.0cm, 0.0cm);}}

\newrobustcmd*{\mythickdasheddottedline}[1]{\tikz{\draw[line width=0.3mm,draw=#1] (-0.22, 0.1cm) -- (-0.13cm, 0.1cm); \draw[line width=0.3mm,draw=#1] (-0.085cm, 0.1cm) -- (-0.055cm, 0.1cm); \draw[line width=0.3mm,draw=#1] (-0.01cm, 0.1cm) -- (0.08cm, 0.1cm); \draw[line width=0.3mm,draw=#1] (-0.0cm, 0.0cm);}}

\newrobustcmd*{\mycircle}[2]{\tikz{\draw[draw=#1,fill=#2] (0,0)
circle (0.1cm);}}

\newrobustcmd*{\mythickcircle}[2]{\tikz{\draw[line width=0.3mm,draw=#1,fill=#2] (0,0)
circle (0.1cm);}}

\newrobustcmd*{\mydot}[1]{\tikz{\draw[line width=0.3mm,draw=#1] (0,0)
circle (0.025cm);}}
\DeclareMathOperator*{\argminA}{arg\,min}

\graphicspath{ {} }

\usepackage{scalerel}
\usepackage{stackengine,wasysym}

\biboptions{comma,sort&compress}
\begin{document}
\begin{frontmatter}

\title{Data-driven Analysis of Relight variability of Jet Fuels induced by Turbulence}
\author[label1]{Malik Hassanaly\corref{cor1}}
\ead{malik.hassanaly@gmail.com}
\author[label2]{Yihao Tang\fnref{anss}}
\author[label2]{Shivam Barwey}
\author[label2]{Venkat Raman}
\address[label1]{National Renewable Energy Laboratory (NREL), Golden, CO 80401, USA}
\address[label2]{Department of Aerospace Engineering, University of Michigan, Ann Arbor, MI 48109, USA}

\cortext[cor1]{Corresponding author:}
\fntext[anss]{Current address: ANSYS, Inc., Lebanon, NH 03766, USA}
%\fntext[nrel]{Currently at }

\begin{abstract}

For safety purposes, reliable reignition of aircraft engines in the event of flame blow-out is a critical requirement. Typically, an external ignition source in the form of a spark is used to achieve a stable flame in the combustor. However, such forced turbulent ignition may not always successfully relight the combustor, mainly because the state of the combustor cannot be precisely determined. Uncertainty in the turbulent flow inside the combustor, inflow conditions, and spark discharge characteristics can lead to variability in sparking outcomes even for nominally identical operating conditions. Prior studies have shown that of all the uncertain parameters, turbulence is often dominant and can drastically alter ignition behavior. For instance, even when different fuels have similar ignition delay times, their ignition behavior in practical systems can be completely different. In practical operating conditions, it is challenging to understand why ignition fails and how much variation in outcomes can be expected. The focus of this work is to understand relight variability induced by turbulence for two different aircraft fuels, namely Jet-A and a variant named C1. A detailed, previously developed simulation approach is used to generate a large number of successful and failed ignition events. Using this data, the cause of misfire is evaluated based on a discriminant analysis that delineates the difference between turbulent initial conditions that lead to ignition or failure. From the discriminant analysis, a compressed sensing algorithm is then applied to help pinpoint the locations of relevant turbulent features. Findings from the discriminant analysis are confirmed with the time history of near kernel properties. Next, a clustering strategy is used to identify ignition and misfire modes. With this approach, it was determined that the cause of ignition failure is different for the two fuels. While it was found that Jet-A is influenced by fuel entrainment, C1 was found to be more sensitive to small scale turbulence features. A larger variability is found in the ignition modes of C1, which can be subject to extreme events induced by kernel breakdown. 

\end{abstract}

\end{frontmatter}

\clearpage
\setcounter{page}{1}
%\tableofcontents
\section{Introduction}

High altitude relight is an important safety requirement for aircraft engines. In the event of a flame blow-out during operation, a robust procedure for reigniting the combustor towards stable power output is a mandatory certification requirement. With the introduction of alternative jet fuels (AJFs), there is a need to determine the robustness of relight with varying fuel properties~\cite{colket2017overview}. Relight involves the injection of a spark kernel into the combustor, followed by the complex processes of kernel propagation, mixing, ignition, and flame stabilization~\cite{mastorakos2009ignition,tang2019numerical}. Since this ignition process is highly sensitive to the flow conditions inside the combustor as well as the spark properties, precisely determining whether a single kernel will ignite, or in the event of ignition, how the ignition occurs, is not possible. Instead, the ignition process should be analyzed statistically using an ensemble of realizations, which is the approach used here.

Of the many sources of uncertainty, turbulence is found to be the dominant factor~\cite{chakraborty2007effects,mastorakos2009ignition,tang2019numerical}. For AJFs in particular, the ignition process in the presence of turbulence can be different compared to laminar ignition properties that depend only on the fuel oxidation chemistry. For instance, three different fuel compositions, namely Jet-A and two variants, C1 and C5, were considered in Ref.~\cite{sforzo2017liquid}. While laminar calculations and ignition delay experiments showed that Jet-A and C1 were similar in ignition characteristics, with C5 showing longer ignition delay times, ignition in turbulent flow was similar between Jet-A and C5, while C1 showed very low ignition probability. These findings were also confirmed using numerical simulations~\cite{yihao_PROCI}. Since fuel oxidation involves a range of time-scales, its interaction with the multi-scale turbulent flow can alter the impact of specific reaction pathways~\cite{aspden2019priori}. Hence, the ability to determine this coupling between turbulence and ignition is essential for ensuring robust relight.

The effect of turbulence on the ignition process has been studied using simulations and experiments in the past~\cite{blouch2003effects,ahmed2007measurements,wawrzak2019spark,chakraborty2007effects,xiong2001investigation,eichenberger1999effect,echekki2007regime,vasudeo2010regime,reddy2011numerical,kolera2006direct,chi2018direct,yoo2011direct}. In premixed flames, fundamentals of vortex-kernel interaction have been investigated both experimentally~\cite{xiong2001investigation,eichenberger1999effect} and numerically~\cite{echekki2007regime,vasudeo2010regime,reddy2011numerical,kolera2006direct}, leading to advances in understanding the cause of misfire and in characterizing ignition success and failure pathways. While vortices can enhance the growth rate of the flame kernel through wrinkling, they can also stretch the flame front and induce local extinction~\cite{xiong2001investigation,kolera2006direct}. Alternatively, a vortex can break through a kernel; in the case of an energetic vortex, global extinction can occur~\cite{echekki2007regime}. In some cases, it was also observed that residuals of the global extinction could later reignite~\cite{vasudeo2010regime}. These detailed analyses of kernel-vortex interaction could only be conducted in a laminar regime where a single vortex interacted with a kernel. Although necessary to guide design, the same analysis is much more complex in a chaotic turbulent flow~\cite{renard2000dynamics}, as many types of vortices interact with the flame kernel, causing all the aforementioned effects to be intertwined. While studies in a more disorganized flow have been recently conducted \cite{chi2018direct,saito2018effects}, the focus was on the effect of turbulence intensity rather than realization-to-realization variations.

In the context of non-premixed ignition, counterflow flames have been studied experimentally showing certain unique features. It was found that in the turbulent regime, the ignition temperature could be lower than in a laminar case, not only due to spark kernel wrinkling but also because the fuel was fed differently to the kernel~\cite{blouch2003effects}. Furthermore, ignition was found to occur even when the kernel was introduced in a region located outside the flammability limits~\cite{ahmed2007measurements}, which suggests that turbulence and mixing far from the igniter (non-local effects) can influence the ignition outcome. Similar non-local effects were recently evidenced through direct numerical simulations of a turbulent mixing layer \cite{wawrzak2019spark}. In isotropic turbulence with inhomogeneous mixing, it was found that turbulence could also have an adverse effect on ignition by accelerating the dissipation of heat prior to establishing the flame front~\cite{chakraborty2007effects}. 

Compared to premixed systems, inhomogeneity of the fuel distribution adds a layer of complexity to the ignition problem, which might explain why the ignition and misfire processes have been mostly described macroscopically. In the case of ignition in a mixing layer, it was found that the flame develops inside the layer and stabilizes along the stoichiometric surface \cite{wawrzak2018implicit}. In turbulent jet flames, numerical simulations have shown that the ignition sequence can be separated into four phases: 1) kernel initiation, 2) quasi-spherical expansion, 3) upstream propagation, 4) stabilization \cite{lacaze2009large,ahmed2006spark}. At the same time, studies have shown that inside the same combustor, ignition and failure can occur in several ways (called hereafter as ignitions and failure modes respectively) \cite{de2019ignition,letty2012structure}, which suggests that a finer description of the ignition sequences can shed light on new ignition and failure mechanisms.

From a design standpoint, the effect of turbulence on ignition outcome raises two main questions addressed in the present work: 1) since turbulence affects the ignition outcome, it is desirable to pinpoint the turbulence structures that induce ignition failure in a realistic combustor; 2) since turbulent ignition and misfire can themselves occur in different ways, knowing whether or not the combustor has ignited is not sufficient to precisely characterize the response of the combustor to the sparking process. Therefore, a method that identifies ignition and failure pathways would be useful. 

Recently, Tang et al.~\cite{tang2019numerical} have recently demonstrated an end-to-end simulation approach for generating kernel ignition and failure events, which can then be used to compute ignition probabilities \cite{yihao_PROCI}. In the present context, this approach directly handles the impact of variations in ignition outcomes due to turbulent initial conditions. In this work, a large ensemble of kernel injection events with either success or failure of ignition is generated using the aforementioned modeling procedure. The resulting data is then used to infer the cause of ignition failure, and identify ignition modes.

In the current work, the design questions discussed above are answered for two different jet fuels, namely Jet-A and C1~\cite{colket2017overview}. The rest of the paper is organized as follows: Section~\ref{sec:config} describes the configuration simulated here as well as the numerical method used to generate the dataset. Section~\ref{sec:dataset} describes the dataset generated using an ensemble of numerical simulations. The results, along with the specific data-driven method used are separated in the following two sections. The inference process that determines the cause of ignition failure is explained in Sec.~\ref{sec:causalityInference}. The ignition and failure modes of both fuels are identified and studied in Sec.~\ref{sec:ignitionModes}. Section~\ref{sec:summary} summarizes the findings and provides conclusions on the viability of the proposed methods.

\section{Simulation configuration and computational details}
\label{sec:config}

\begin{figure}[ht]
\centering
\includegraphics[width=0.95\textwidth]{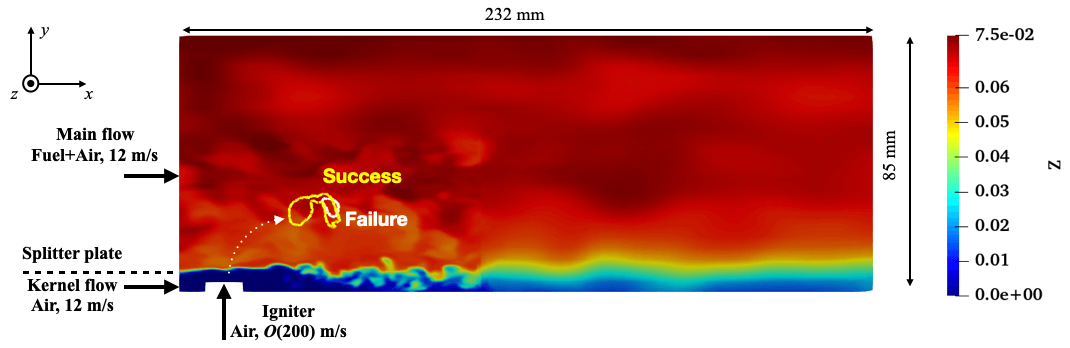}

\caption{Schematic of the simulated configuration illustrated with a mixture fraction contour at the mid-plane in the spanwise direction. The trajectory of a kernel is sketched by the dashed white line. It has two possible outcomes: success or failure. The successful and failing ignition contours correspond to the isocontour of $C=0.0035$ after $2.8~ms$ for a successful and failing ignition of the C1 fuel.}
\label{fig:config}
\end{figure}

The simulated configuration (Fig.~\ref{fig:config}) features non-premixed forced ignition with prominent strain and non-local effects~\cite{sforzo2017liquid,sforzo2015ignition}. The flow field is characterized by a mixing layer between the upper and lower streams. The upper main flow contains a fuel-air mixture with appreciable spatio-temporal equivalence ratio variations. The lower kernel flow issues air at the same bulk velocity as the main flow. The ignition kernel is introduced using an igniter immersed into the bottom wall of the domain. The ignition may succeed or fail depending on both the kernel properties (e.g. energy, velocity, etc.) and turbulence properties (fuel mixing and strain) \cite{sforzo2014thesis}.

Here, two realistic jet fuels (Jet-A and C1) are simulated. The operating conditions (Tab.~\ref{t:operating_condition}) are adopted from Ref.~\cite{sforzo2017liquid}, but include adjustments in the kernel energy to ensure an ignition probability close to $50\%$.
\begin{table}[t]
\begin{center}
\setlength{\abovecaptionskip}{0pt}
\setlength{\belowcaptionskip}{5pt}
\caption{Operating conditions}
\label{t:operating_condition}
\begin{tabular}{lcl}
\hline
Properties           & \multicolumn{2}{c}{Value}            \\ \hline
Fuel/oxidizer        & Jet-A/air                  & C1/air  \\
Equivalence ratio    & 1.1                        & 1.1     \\
Spark energy ($E_d$) & 0.96 J                     & 1.1275 J \\
Inflow temperature   & 475 K                      & 475 K   \\
Bulk inflow velocity      & \multicolumn{1}{l}{12 m/s} & 12 m/s \\
Total simulation time     & 3.5 ms & 3.5 ms \\ \hline

\end{tabular}
\end{center}
\vspace{0mm} 
\end{table}
 
 The ignition simulation database is created using the framework developed by Tang et al.~\cite{yihao_cnf,yihao_asme} and is only briefly described here. The forced ignition is simulated using large-eddy simulations (LES) with tabulated chemistry, where the look-up table provides the source term of progress variable to the LES solver based on the local filtered $\{H, Z, C\}$, where $H$ is the total enthalpy, $Z$ is the mixture fraction and $C$ is the progress variable defined as $C=Y_{H2O}+Y_{H2}+Y_{CO}+Y_{CO2}$. 
% This next paragraph can be cut down
The model uses a hybrid tabulation strategy that automatically adapts between a homogeneous reactor model (HR) and a flamelet/progress variable approach (FPVA) depending on the stages of the ignition process. The spark kernel is modeled as an energy deposition (ED) of high $H$, which corresponds to the HR region in the tabulation. The reaction rate obtained from HR calculations allows for the initiation of reactions even starting from a zero $C$. As the kernel moves through the flow field, entrainment of cold outer flow reduces $H$, thereby transitioning to the FPVA region in the tabulation. Additional details of the hybrid tabulation construction are available in~\cite{yihao_cnf}. The tabulation for the different fuels is obtained based on the HyChem family of chemistry mechanisms~\cite{wang2018physics_1,xu2018physics_2,wang2018physics_4,gao2017reduced}. In another article, the applicability of this framework to Jet-A, C1, and C5 fuels has been validated against experimental data~\cite{yihao_PROCI}. It has been demonstrated that the approach accurately predicts the ignition trends seen in the experiments.

The sparking effects are modeled by introducing a high-energy pulsing jet that emerges from the igniter top surface. The velocity and enthalpy profiles enforced at the igniter boundary are chosen to match the kernel size and trajectory measurements~\cite{sforzo2017liquid}. The enforced enthalpy profile is dependent on the spark energy deposited $E_d$. To ensure that the number of ignition/failure samples are approximately equal, the spark energy ($E_d$) is set to the values shown in Tab.~\ref{t:operating_condition}. The precise boundary conditions are set in the same manner as in Ref.~\cite{yihao_cnf,yihao_PROCI}.%{\color{red} Yihao, please explain a bit more the boundary conditions}

\begin{figure}[ht]
\centering
\includegraphics[width=0.95\textwidth]{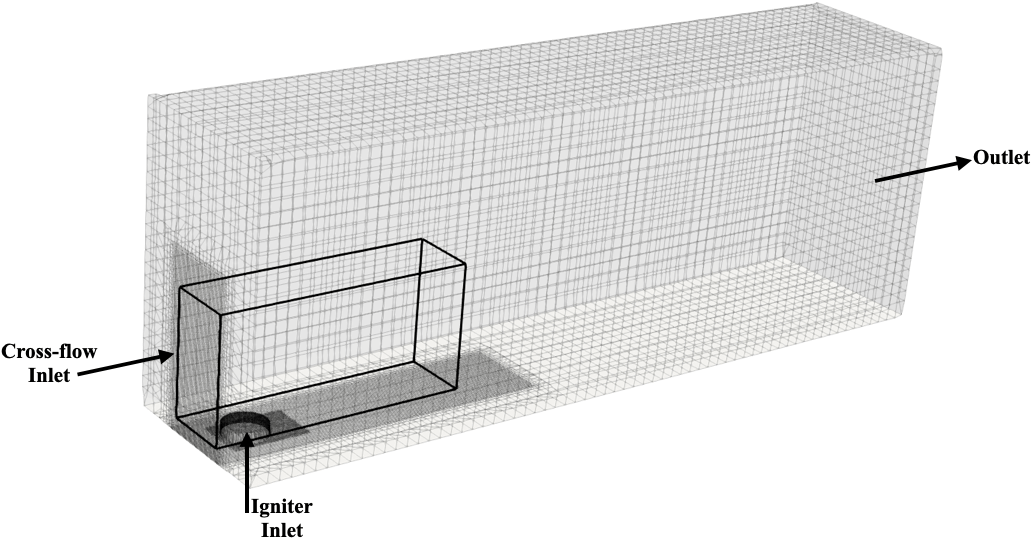}
\caption{Mesh used for the LES. The black box denote the computational domain used for the data-driven analysis.}
\label{fig:mesh}
\end{figure}

In the current work, the variability in ignition is attributed solely to the turbulent initial conditions. These initial conditions are obtained by sampling a cold flow simulation every $t_{samp}=5$~ms run over $2.5$~s. The sampling rate is chosen such that 

\begin{equation}
    t_{samp} > l_{int}/U_{bulk},
\end{equation}

where $l_{int}$ is the integral length scale estimated as the ensemble and spatial average of $\frac{k^{3/2}}{\varepsilon}$, where $k$ is the turbulent kinetic energy, $\varepsilon$ is the turbulent dissipation rate estimated from the filtered strain rate tensor, $U_{bulk}$ is the bulk streamwise velocity, here $12$~m/s. For C1, $l_{int}/U_{bulk} \approx 3$~ms and for Jet-A, $l_{int}/U_{bulk} \approx 2.9$~ms. The choice of $t_{samp}$ ensures that the sampled initial conditions are independent. In total, $500$ reacting flow simulations are performed for each fuel, and are advanced for $3.5$~ms, which was found sufficient to determine whether a simulation leads to ignition or failure (see Sec.~\ref{sec:dataset} for the precise definition of ignition and failure). The 3D domain is discretized with a hexahedron-dominant $2$~million cell mesh (shown in Fig.~\ref{fig:mesh}). The near-igniter region is discretized with a cell size of $0.2$~mm (about 20 points across the igniter), and the maximal cell size in the volume traversed by the kernel during the $3.5$~ms is $0.4$~mm. The far-field is discretized with a cell size of $3.2$~mm. The mesh is illustrated in Fig.~\ref{fig:mesh}.

The computations are carried using a variable density low-Mach solver~\cite{malik-openfoampaper} that is based on the OpenFOAM framework. Each simulation was run on 252 processors and took on average one hour to reach completion. The total computational cost approached $0.3~$million CPUh. The non-linear closure terms for sub-filter transport are modeled using a dynamic subgrid-scale model~\cite{germano}. The turbulent diffusivity is obtained using a constant turbulent Schmidt number ${Sc}_t = 0.72$. A constant turbulent Prandtl number ${Pr}_t = 0.7$ is used for the energy equation. The subfilter probability density function (PDF) of mixture fraction is assumed to follow a $\beta$-PDF with variance $Z_{var}$ defined as an algebraic relation using the filtered $Z$~\cite{pierce1998dynamic}. The PDFs of $C$ and $H$ are assumed to be described by $\delta$-PDF. The chemistry is tabulated as a function of four variables, $\{C,Z,H,Z_{var}\}$.

\section{Dataset}

\label{sec:dataset}
LES computations are used to characterize individual ignition realizations. Although one LES should, in general, reproduce the evolution of an ensemble of realizations~\cite{langford-moser}, it is acceptable here to consider that each LES is representative of a single realization. Given that each simulation runs for a short period of time, it can be expected that the LES closely tracks any DNS realization that corresponds to a particular initial condition. Additional justification is provided in Appendix~\ref{app:indivReal}. The 3D results are sampled $T=21$ times per simulation between $0$ and $3.5$~ms with points clustered at early times. To ease the manipulation of data, the unstructured field is interpolated onto a structured grid made of $M = 160 \times 100 \times 40$ grid points with uniform spacing in a domain surrounding the igniter, given by $74.5~mm~\times~36.5~mm~\times~20~mm$ in the x, y, and z-directions respectively (see black box in Fig.~\ref{fig:mesh}). For each fuel, the whole dataset comprises an ensemble of scalar fields ${\bf x}_i^{t}\in \mathbb{R}^{M}$, where $t = 1,\ldots,T$ and $i = 1,\ldots,N$ ($N$ being the number of LES realizations).

\begin{figure}[ht!]
\centering
% Low res
\includegraphics[width=0.45\textwidth]{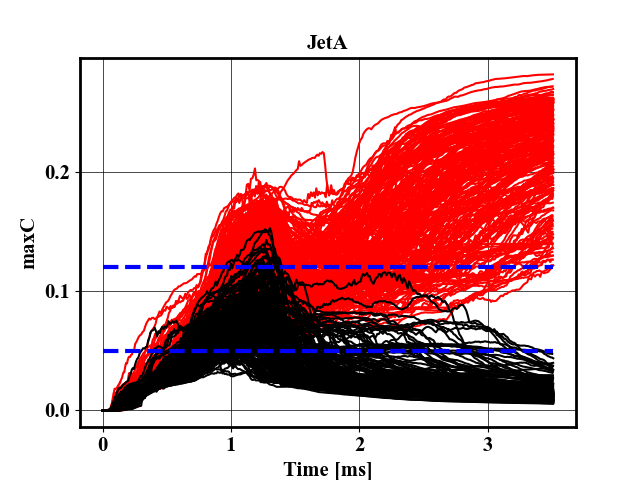}
\includegraphics[width=0.45\textwidth]{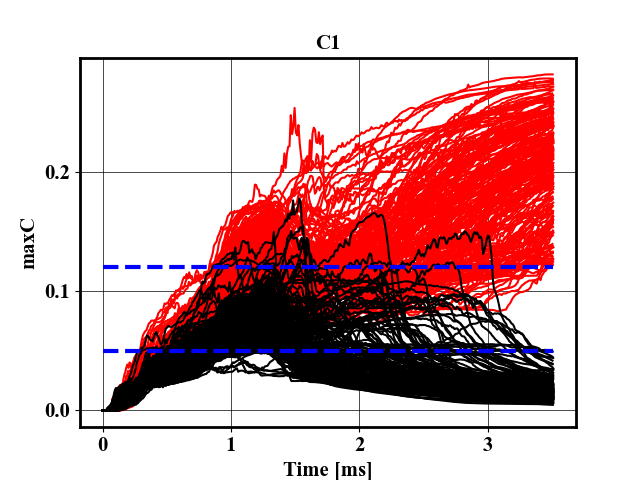}

% High res
%\includegraphics[width=0.45\textwidth]{maxC_JetA.eps}
%\includegraphics[width=0.45\textwidth]{maxC_C1.eps}

\caption{Time history of maximal value of progress variable over the computational domain for realizations labelled as ignition (\mythickline{red}) and misfires (\mythickline{black}), for the 467 realizations of Jet-A (left) and 432 realizations of C1 (right), overlaid with the bounds that define ignition (0.12) and failure (0.05) (\mythickline{blue}).}
\label{fig:criterion}
\end{figure}

The ignition outcome (success or failure) is determined using the maximum value of the progress variable field at the final simulation time $t = T_f = 3.5$~ms. At the final time, if the progress variable value exceeds anywhere in the domain 0.12, the ignition is considered successful. In turn, if the progress variable value does not exceed 0.05 anywhere, the ignition is considered failed. It was found that only in a few simulations lead, at $t=T_f$ to a maximum value of progress variable between 0.05 and 0.12. Formally, 
\begin{equation}
\begin{split}
\max{({\bf c}_i^{t=T_f})} > 0.12 \implies \forall t, {\bf x}_{i}^{t} \in {\cal S}, \\
\max{({\bf c}_i^{t=T_f})} < 0.05 \implies \forall t, {\bf x}_{i}^{t} \in {\cal F}, 
\end{split}
\end{equation}
where $\cal S$ and $\cal F$ are the sets of ${\bf x}_i$ containing respectively ignition successes and failures, and ${\bf c}_i^{t=T_f}$ is the final progress variable field for the $i$-th LES. For Jet-A, $\cal S$ and $\cal F$ have sizes of $237$ and $230$, respectively; for C1, the corresponding amounts are $212$ and $220$. For both fuels, $N = \text{card}(\cal S) + \text{card}(\cal F)$. Figure~\ref{fig:criterion} shows the time history of the maximal value of progress variable for each realization and each fuel, and highlights that the criterion clearly demarcates ignition (continuously increasing curves) and misfires (continuously decreasing curves). Some of the simulations had a final maximal progress variable value that fell in the range [0.05, 0.12]. In that range, it was found that the maximal value of progress variable was neither continuously increasing or decreasing, which prevented labeling the realization as ignition success or failure. It is likely that these simulations correspond to a late ignition or late failure which happens after $3.5$~ms. To avoid polluting the dataset with false success or false failure, these simulations were discarded.

\section{Cause of ignition failure}
\label{sec:causalityInference}

%{\color{red} This section (4) and section 5 seem to provide very similar outputs and you use two different methods. This text structure is very confusing. You have to explain clearly a) what is meant by cause of ignition failure, and b) what are ignition failure modes. At the end of the introduction, this should be made clear. When you start section 4, explain again what this is, and again at start of section 5}

The effect of vortices on ignition outcome has mostly received attention in well-organized flows, where a single vortex interacts with a flame kernel \cite{xiong2001investigation,eichenberger1999effect,echekki2007regime,vasudeo2010regime,reddy2011numerical,kolera2006direct}. These studies showed that only vortices of a certain size and strength can induce misfire. While such canonical studies are useful to understand the fundamentals of vortex-kernel interaction, a different approach is needed for realistic combustors where many vortices interact with the spark kernel for an extended period of time.

Using the generated data, the first objective is to utilize a method that can extract the turbulence structures which contribute the most to the ignition outcome, i.e. a method that provides the cause of ignition failure. Here, ignition failure occurs not because of a change in operating conditions, but because of initial condition variability. The objective is to infer what turbulent structures in the initial conditions affect the ignition outcome. The causality inference is done using two main methods that are described in the following subsections. The physical interpretation is then provided.

\subsection{Post-processing methodology}
\label{sec:postProcessing}

Two methods are used to pinpoint what turbulence features induce misfires. To identify how the initial conditions differ between igniting and failing cases, a discriminant analysis combined with a compressed sensing approach on the mixture fraction fields is used (Sec.~\ref{sec:lda}). At later times, the conditional averages of kernel surface properties are used to understand the misfire process (Sec.~\ref{sec:kernelSurf}).

\subsubsection{Discriminant analysis}
\label{sec:lda}

% Describe main idea: want to figure out the the difference between success and failure.
% first define the goal and why we cannot look at snapshots individually
% Then describe the method in general terms: data -> pod -> lda -> sensing. need a new figure. 
The goal of this section is to find the spatial locations that are most relevant for identifying the igniting and failing cases. This objective is reformulated as follows. Each snapshot in $\cal S$ and $\cal F$ can be considered as a member of an $M$ degree-of-freedom (DOF) phase space, where $M$ is the number of grid points (or pixels) present in a single snapshot. The goal is to determine a small subset of these $M$ dimensions or grid points that plays an important role in distinguishing whether or not an initial condition will lead to ignition, and to potentially connect this subset to turbulence-based causal features. Ultimately, this identified subset of grid points should identify locations in physical space related to high variability in the ignition outcome.

To accomplish this goal, one route is to individually inspect each snapshot and extract key features that delineate ignition success and failure using expert-guided analysis. However, since $M$ (as well as $N$, the number of realizations) can be very large, this approach is infeasible. Instead, a more practical route is to employ data-driven techniques that take advantage of the fact that the ignition outcome for each initial condition to be analyzed is already known. As such, the technique employed here (and to be described in further detail below) draws from a method based on supervised classification and sparse sensing \cite{brunton2016sparse,bai2017data}. The method relies on a combination of proper orthogonal decomposition (POD, \cite{sirovich1987}) and linear discriminant analysis (LDA, \cite{lda_book}) to eventually produce the desired subset of grid points, termed \textit{sensors}, of which there are $N_S$. The term "sparse" here refers to the fact that the method ensures $N_S \ll M$.

\begin{figure}
    \centering
    \includegraphics[width=0.8\textwidth]{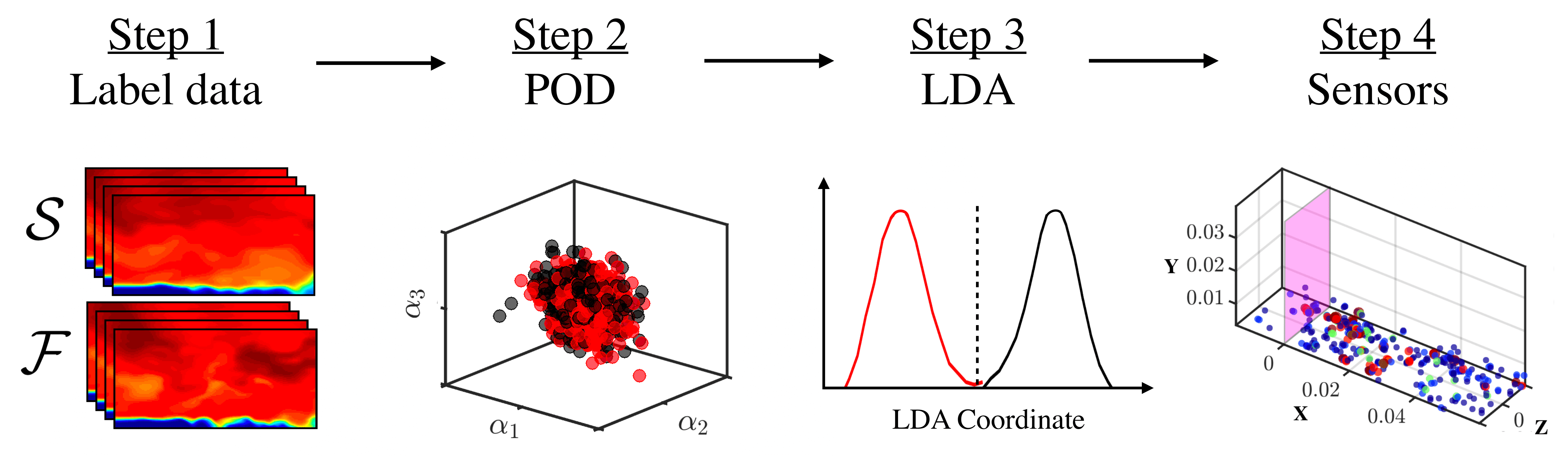}
    \caption{Illustration of the sparse sensing workflow. In POD and LDA representations, red is ignition success and black is failure. No clear separation is seen in first three POD coordinates, but the densities show clear separation in the 1-dimensional LDA space, which is the desired outcome.}
    \label{fig:lda_schematic}
\end{figure}

The workflow of the sparse sensing technique can be categorized into four steps. In step 1, the data in question is organized into classes of interest, here $\cal S$ and $\cal F$ (as described in Sec.~\ref{sec:dataset}). In step 2, in the usual case that $M \gg N$, the data is pre-processed using a variant of POD known as the method of snapshots \cite{sirovich1987}. As a result of the second step, the data is transformed into a so-called POD-space of $K$ dimensions, where $K \leq N$ is the number of retained POD modes. In step 3, LDA is carried out in the POD-space. The \textit{LDA vector} is obtained as a result of this third step, which can be interpreted as a visualizable ``flow direction" in physical space (not dissimilar to a POD mode) that discriminates the classes of interest. Then, in step 4, an optimization problem is solved to produce a sparse counterpart to the LDA vector, which provides the desired $N_S$ sensor locations. Inspection of these sensors in physical space then facilitates the analysis of turbulence-driven ignition misfires. This workflow is summarized in Fig.~\ref{fig:lda_schematic}.

Additional detail on the LDA procedure (step 3) is provided below, followed by details on the recovery of the sensors from the LDA output (step 4). It is assumed that the reader is familiar with the basics of POD (step 2), the methodology of which is described in Ref.~\cite{sirovich1987}.

% Original paragraph: 
% When analyzing spatiotemporal features over a large set of data, it is infeasible to parse through each snapshot and extract delineating features. This is mainly because each snapshot can be considered as a $M$ degree of freedom (DOF) data, where $M$ is the number of grid points or pixels (in the case of experiments) present in a single snapshot. Typically $M$ is large, and the goal is to determine the subset that shows the maximum variability in the outcome. The discriminant analysis combined with compressed sensing~\cite{bai2017data} can help isolate these spatial locations. Here, the objective is to find some sensors $N_S$ in the physical domain whose locations identify the key delineating features of ignition success and failure, with the additional requirement that $N_S \ll M$. It will be seen below that the sensors can also be used to ensure that the discriminant is not the result of overfitting the data. The approach used to obtain the sensor locations draws from the work of Ref.~\cite{brunton2016sparse,bai2017data}. It is based on a combination of proper orthogonal decomposition (POD)~\cite{sirovich1987} and linear discriminant analysis (LDA)~\cite{lda_book}. Below, LDA is briefly described and is then connected to the sparse sensing goal.

LDA is a supervised classification technique that is used primarily as a tool to label some quantity to one of several classes using a trained ``discriminator". In the two-class problem considered here, the inputs to the algorithm are the sets of pre-labeled data (here $\cal S$ and $\cal F$), and the output is a single discrimination vector, $\bf w$, termed the LDA vector. The LDA vector is by design an optimal linear discriminator in the sense that a projection of the data in $\cal S$ and $\cal F$ onto {\bf w} generates the \textit{LDA coordinates} that attempt to maximize between-class distance and preserve within-class variance in the LDA coordinate system. There also exist more complex discriminant analysis techniques such as non-linear discriminant analysis (NLDA). In NLDA, the data is transformed with a non-linear transformation and the LDA is applied to the transformed data \cite{roth2000nonlinear}. It can be useful to turn to non-linear techniques if the LDA is unsuccessful. Here, the LDA alone was found sufficient to distinguish igniting and failing initial conditions.

As mentioned previously, for computational ease, the data is first transformed into a $K$-dimensional POD representation via the method of snapshots~\cite{sirovich1987}, where $K$ is the number of retained modes (step 2 in Fig.~\ref{fig:lda_schematic}). Since $K \ll M$, this allows for significant computational speed-up in LDA execution. The K-dimensional LDA vector is obtained by maximizing the objective function
\begin{equation}
    {\cal L({\bf w})} = \frac{{\bf w}^T C_{B} {\bf w}} {{\bf w}^T C_W {\bf w}},
\end{equation}
where the $K$-by-$K$ matrices $C_B$ and $C_W$ represent the between-class and within-class covariances, respectively. These are computed using the class-conditional averages, i.e. 
\begin{equation}
    C_B = (\boldsymbol{\mu}_{\cal S} - \boldsymbol{\mu}_{\cal F})(\boldsymbol{\mu}_{\cal S} - {\boldsymbol{\mu}}_{\cal F})^T, 
\end{equation}
\begin{align}
    C_W &= {\langle ({\bf a} - \boldsymbol{\mu}_{\cal S})({\bf a} - \boldsymbol{\mu}_{\cal S})^T \mid {\bf a}_i \in S \rangle} \\
    & + {\langle ({\bf a} - \boldsymbol{\mu}_{\cal F})({\bf a} - \boldsymbol{\mu}_{\cal F})^T \mid {\bf a}_i \in F \rangle}. \nonumber
\end{align}
Here, $\boldsymbol{\mu}_{\cal S} = \langle {\bf a} \mid {\bf a}_i \in {\cal S} \rangle$ and $\boldsymbol{\mu}_{\cal F} = \langle {\bf a} \mid {\bf a}_i \in {\cal F} \rangle$. It can be shown that the LDA vector $\bf w$ is the eigenvector corresponding to the maximum eigenvalue of $C_W^{-1} C_B$~\cite{brunton2016sparse, lda_book}. 

An illustration of effective LDA discrimination is shown in Fig.~\ref{fig:lda_schematic}. Note that the POD coefficients obtained from the individual classes of data may not be distinguishable in the original phase space. If an LDA coordinate, which is a linear combination of POD coefficients, can be found, then the data can be demarcated, as shown in Fig.~\ref{fig:lda_schematic}. As will be discussed below, even the inability of the LDA vector to demarcate the classes (i.e., the PDFs are overlapping in Fig.~\ref{fig:lda_schematic}) is an important piece of information that could be used to understand the causal mechanism.

% Illustration of an ideal LDA process using mixture fraction fields. In POD and LDA representations, {\color{red} red is ignition success and black is failure}. No clear separation is seen in the first three POD coordinates, but the densities show clear separation in the 1-dimensional LDA space, which is the desired outcome.

The sparse sensors are then computed directly from the LDA vector. The representation of the LDA vector in the original $M$-dimensional phase space, denoted ${\bf s} \in \mathbb{R}^M$, can be obtained using the POD modes (i.e. ${\bf s} = \phi {\bf w}$, where $\phi \in \mathbb{R}^{M \times K}$ is the POD basis). In this sense, ${\bf s}$ represents a single, visualizable flow direction that discriminates the two classes of interest defined in the LDA problem with respect to the scalar field used to generate the LDA vector. The sparse vector ${\bf \hat{s}} \in \mathbb{R}^M$ is obtained as the solution to the following optimization problem: 
\begin{equation}
    \label{eq:optim}
    \min_{\hat{\bf s}} \| \hat{\bf s} \|_1 \quad \text{s.t.} \quad \phi^T \hat{\bf s} = {\bf w}.
\end{equation}
Upon convergence of the optimization, the number of sensors $N_S$ is the number of non-zero elements of the vector ${\bf \hat{s}}$. Note that both ${\bf \hat{s}}$ and ${\bf s}$ are similar in that they 1) exist on the original $M$-dimensional grid, and 2) produce the same LDA vector ${\bf w}$ in the POD space. Thus, assuming good convergence in Eq.~\ref{eq:optim}, analysis of the $N_S$ sensors (the non-zero pixels in ${\bf \hat{s}}$) provides a low-dimensional representation of the discriminative features between the two classes used to generate the LDA vector ${\bf w}$. To elucidate the meaning of the LDA vector and sensors, the LDA procedure is demonstrated on a toy problem in Appendix~\ref{app:toy_lda} -- examples of an LDA vector and sensors obtained from the mixture fraction dataset (${\bf s}$ and ${\bf \hat{s}}$, respectively), are also shown.

As will be seen in Sec.~\ref{sec:causeResults}, there is no guarantee for the LDA vector to be interpretable since it can overfit the input data. It is important to assess classification accuracy (termed LDA accuracy) on the dataset used to generate the LDA vector itself. LDA classification is performed by computing the distance between a snapshot and the respective means $\boldsymbol{\mu}_{\cal S}$ and $\boldsymbol{\mu}_{\cal F}$ in the ``LDA space", i.e. after projection onto the LDA vector $\bf w$. A snapshot closer to the conditional mean of one class in the LDA space is assigned to that class. The LDA accuracy represents the ability of the LDA to recognize igniting and non-igniting initial conditions. To further ensure no overfitting, the sparse version of the LDA accuracy can be obtained using the sparse representation of the LDA vector, $\hat{\bf s}$. This step is achieved by generating a dataset using an element-wise multiplication of each snapshot ${\bf x}_i$ with the sensor locations in ${\bf \hat{s}}$ (only the data corresponding to the location of the sensors is kept, with zeros everywhere else). The LDA classification accuracy in terms of this dataset then provides a measure of the overall viability and utility of the recovered sensors. If the sparse classification accuracy is reasonably high, it can be concluded that the discrimination power of the LDA is not a result of overfitting, and thus the sensors can be physically interpreted. On the other hand, if the dataset provides poor classification accuracy, the corresponding sensors should not be considered as meaningful.

As a final measure to address overfitting, a cross-validation training strategy is used when computing the sensor outputs. More specifically, a statistical representation of the sensor locations is obtained by running multiple realizations of the optimization (Eq.~\ref{eq:optim}) using different, randomly chosen subsets from the same input dataset. A single subset, which constitutes a training set for one realization, is obtained by removing 15 snapshots at random from both $\cal S$ and $\cal F$ before computing the LDA vector and the subsequent sensors. The 30 removed snapshots then become the testing set for that same realization, and the LDA accuracies are evaluated on this testing set. The ensemble of optimization runs allows for a more robust analysis of sensor behavior -- in particular, the frequency at which a single grid point is chosen as a sensor can be used to assess sensor importance. Here, 50 realizations of the optimization (i.e. 50 different LDA and sensor outputs) are used to assess the sparse sensing results for each fuel.

\subsubsection{Near-kernel conditional averages}
\label{sec:kernelSurf}

The analysis of initial conditions done with the discriminant analysis isolates turbulent features that differ between initial conditions that lead to ignition or failure. Because that analysis was performed at initial times, it allowed drawing conclusions about the cause of ignition or failure. To confirm those findings, it is informative to track the evolution of the spark kernel at later times to observe if the behavior of the kernel is consistent with the discriminant analysis. Conditional averages among the igniting and failing kernels are useful for this matter. 

Since the focus is on the turbulence-kernel interaction, only quantities at the kernel boundary are extracted. 
The kernel boundary is chosen as the isosurface $C=0.02$, which was found to be indicative of the early stages of kernel development. Note that the results below, however, are broadly insensitive to the choice of isosurface value. The surface is extracted using a standard marching cube method \cite{lorensen1987marching}. Different turbulence properties (namely vorticity and strain rate) are then extracted and interpolated at the kernel isosurface.

At every time instance, the isosurface area can vary from realization to realization. To avoid biasing the results towards early ignition cases which have a larger kernel surface area, the statistics of the quantity $\xi$ plotted are:
\begin{equation}
\begin{split}
\widehat{\xi_{\mathcal{S}}}(t) = E_{\mathcal{S}}( E_{kernel_i | Outcome=\mathcal{S}}(\xi) ), \\
\widehat{\xi_{\mathcal{F}}}(t) = E_{\mathcal{F}}( E_{kernel_i | Outcome=\mathcal{F}}(\xi) ), 
\end{split}
\end{equation}
where $\widehat{\xi_{\mathcal{S}}}(t)$ (respectively $\widehat{\xi_{\mathcal{F}}}(t)$) denotes the near-kernel statistics for successful (respectively failing) ignition, and $E_{kernel_i | Outcome=\mathcal{S}}(\xi)$ (respectively $E_{kernel_i | Outcome=\mathcal{F}}(\xi)$) denotes the estimate average over the kernel surface area for each successful (respectively failing) case. Note that $\widehat{\xi_{\mathcal{S}}}(t)$ (respectively $\widehat{\xi_{\mathcal{F}}}(t)$) is a time-varying statistic of the quantity $\xi$ for all successful (respectively failing) cases. Variances used to quantify uncertainties for $\widehat{\xi_{\mathcal{S}}}(t)$ and $\widehat{\xi_{\mathcal{F}}}(t)$ are computed in a similar manner.

% Average variances used to quantify uncertainties for $\widehat{\xi_{\mathcal{S}}}(t)$ and $\widehat{\xi_{\mathcal{F}}}(t)$ are computed in a similar manner.

\subsection{Results}
\label{sec:causeResults}

\subsubsection{Influential initial conditions}

\begin{figure}[h]
    \centering
    \includegraphics[width=0.6\textwidth]{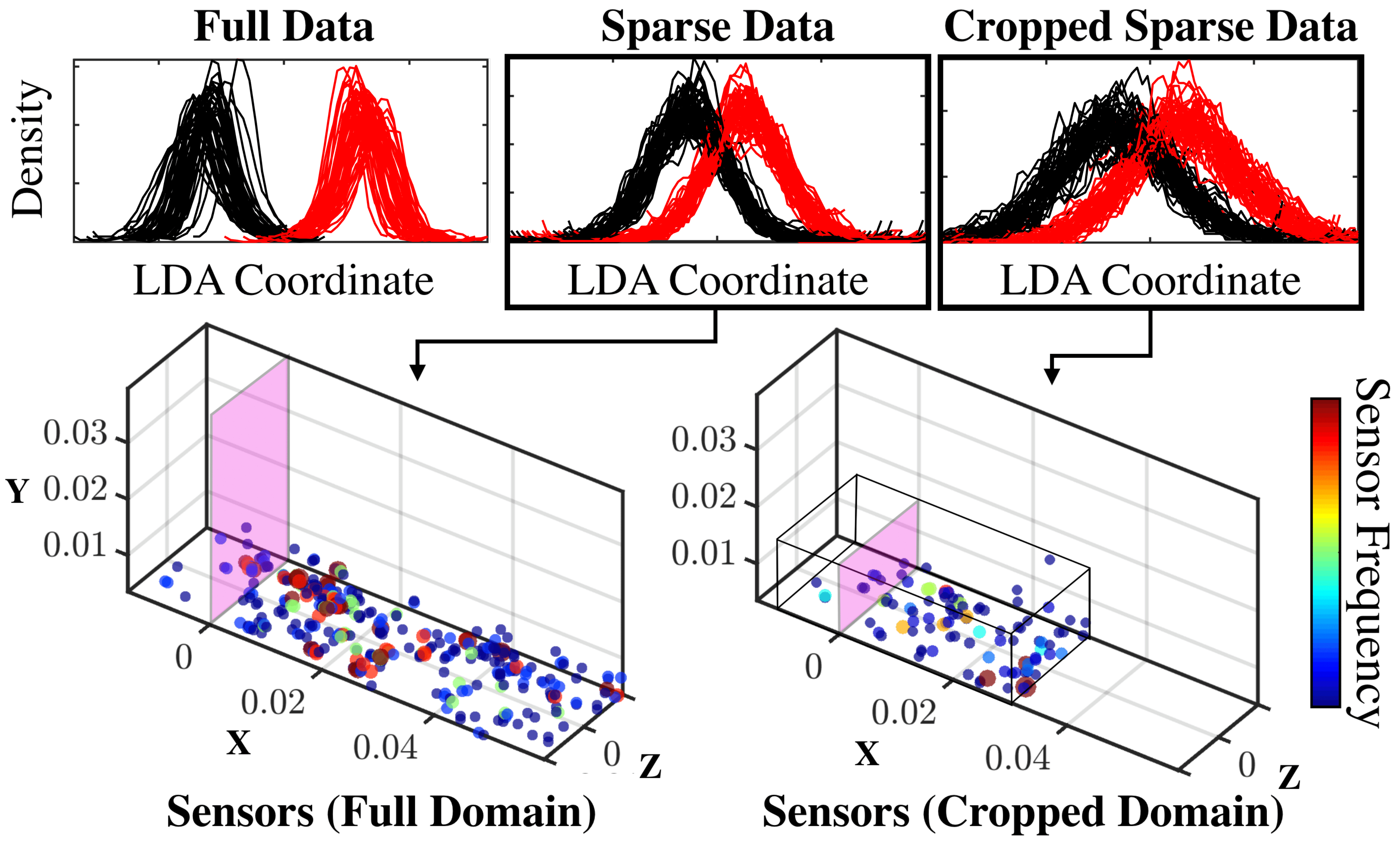}
    \caption{Top: density of mixture fraction at $t=0$ for Jet-A in the LDA space generated by 50 different runs for ignition success (\mythickline{red}) and failure (\mythickline{black}). Left corresponds to original data, middle to sparse data, and right to sparse data computed from runs with a cropped domain. Bottom: corresponding sensor locations with colors indicating the sensor frequency (higher frequency means more probable). Highlighted plane indicates x=0 and box enclosing indicates cropped domain bounds.}
    \label{fig:jeta_Z_sensors}
\end{figure}

The LDA densities and sensor outputs corresponding to the mixture fraction for Jet-A fuel at $t=0$ (the initial condition) are shown in Fig.~\ref{fig:jeta_Z_sensors}. The LDA was performed using $K=400$ POD modes, which retained 99\% of the data variance. In the top row of Fig.~\ref{fig:jeta_Z_sensors}, three LDA densities are shown for three different representations of the input mixture fraction fields. Each LDA density plot contains the results from 50 realizations of the optimization in Eq.~\ref{eq:optim}. The leftmost density plot corresponds to the projection of the full mixture fraction dataset onto the LDA vector ${\bf w}$. The middle plot corresponds to the projection of the sparse mixture fraction dataset (obtained from the sensor locations as described in Sec.~\ref{sec:lda}) onto the same LDA vector. The rightmost density plot corresponds to another sparse dataset for a smaller, cropped domain that is more localized around the igniter -- its purpose is described further below. 

The sensors obtained from the Jet-A mixture fraction fields achieve overall good discriminative power, and the separation in class densities, though not perfectly ideal, is quite pronounced in the sparse datasets. This is a powerful result in itself, as the average number of sensors over the 50 optimizations for the full and cropped domains is $N_S=299.6$ and $332.4$, respectively, which is three orders of magnitude smaller than the original field dimensionality in both cases. In other words, only a few mixture fraction locations at the initial condition are needed to discriminate between ignition success and failure to reasonable accuracy in the Jet-A case. 

Examination of the full-domain sensor locations (bottom-left plot of Fig.~\ref{fig:jeta_Z_sensors}) shows that most high-frequency sensors (i.e. sensor locations that were chosen by many of the optimization runs) are predominantly a) contained in the mixing layer, and b) located to the right of the igniter in the x-direction. 

Note also that some of the sensors in the full-domain case are located far downstream of the igniter and are therefore non-physical. This behavior is a limitation of this method and requires physical insight to rule out some of the sensors. The optimization was also run on a cropped domain (bottom-right of Fig.~\ref{fig:jeta_Z_sensors}) intended to explicitly enforce the sensor locations to coincide with physical, prior expectations. The bounds were chosen to enclose a subspace of the original domain closer to the igniter to eliminate potential downstream noise effects on the sparse sensing output. This can be considered as a user-guided weighting of the near-igniter region. Figure~\ref{fig:jeta_Z_sensors} shows that the high-probability sensors in the cropped domain are also predominantly downstream of the igniter, indicating that this region is indeed influential with regards to Jet-A ignition outcome.

Prior investigations of the present configuration have already evidenced the influence of the time taken by the kernel to reach the mixing layer (transit time) on ignition~\cite{sforzo2014thesis}. 
It is therefore unsurprising to find sensors located at the mixing layer. What was not well-established is that the fuel-kernel mixing above the mixing layer is of little importance. In other terms, the non-local effects are solely contained in the mixing layer, and the early stages of the kernel ejection are the most influential on the ignition outcome. Similar observations were made for different geometries in Refs.~\cite{mastorakos2009ignition,esclapez2015ignition,ahmed2006spark,barre2014flame}. The importance of this downstream part of the mixing layer on ignition is more surprising, as one would instead expect turbulent structures advected from upstream to be more influential. Further interpretation is provided below.

\begin{figure}
    \centering
    % High res
    %\includegraphics[width=0.45\textwidth]{heightLinex.eps}
    % Low res
    \includegraphics[width=0.45\textwidth]{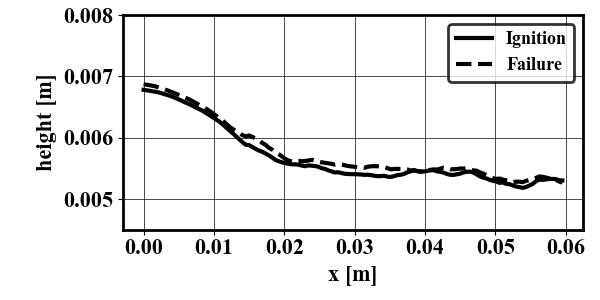}
    % High res
    %\includegraphics[width=0.45\textwidth]{heightLinez.eps}
    % Low res
    \includegraphics[width=0.45\textwidth]{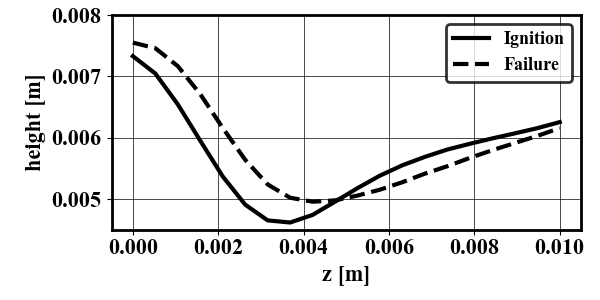}
    \caption{Mixing layer height (in meters) for failing (\mythickdashedline{black}) and igniting (\mythickline{black}) initial conditions, averaged over realizations and the spanwise direction (left)  and streamwise direction (right).}
    \label{fig:height}
\end{figure}

The sparse representation of the LDA vector achieves good discrimination accuracy only if a small collection of points in the original domain can be used to recover a reasonably high class-discrimination. This is possible when localized turbulent structures - indicative of large scales - affect the ignition outcome. It can then be concluded that large structures in the distribution of fuel affect the ignition outcome of Jet-A. The sensors do not provide a physical interpretation by themselves but indicate where the flow should be examined. The sensors shown in Fig.~\ref{fig:jeta_Z_sensors} suggest that the mixture fraction field should be inspected along with the streamwise and spanwise directions. To this end, the mixing layer height is extracted as the height of the isosurface $Z=0.03$. The height is averaged over the igniting and failing initial conditions for Jet-A, and over the streamwise direction, and is shown in Fig.~\ref{fig:height}.

Figure~\ref{fig:height} shows that the mixing layer is the highest near the centerline which is where the igniter is located. Since the mixing layer height decreases with the streamwise direction, it could explain why the downstream region is instrumental for Jet-A ignition. The main difference between the igniting and failing cases can be seen along the spanwise direction where the mixing layer experiences a steep drop, especially in the igniting cases. This phenomenon may be the result of a stronger recirculation caused by the igniter, which will in turn aid the entrainment of fuel in the kernel.

\begin{figure}
    \centering
    % left bot right top
    % high res
    %\includegraphics[width=0.6\textwidth,trim={0cm 0cm 2cm 0cm},clip]{vorticityFilt.eps}
    % low res
    \includegraphics[width=0.6\textwidth,trim={0cm 0cm 2cm 0cm},clip]{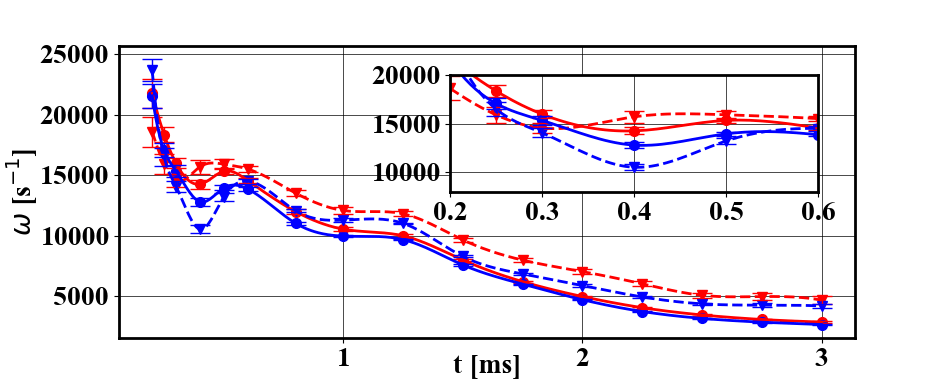}
    \caption{Time evolution of the mean vorticity magnitude at the isosurface $C=0.02$ for Jet-A (blue) and C1 (red) and averaged over the igniting cases  (\mythickErrorcircle{black}{black}) and the non-igniting cases (\mythickErrorVtriangle{black}{black}). Interpolated lines between data points for igniting (\mythickline{black}) and failing cases (\mythickdashedline{black}) are shown to ease visualization. Inset represents zooms on region of interest.}
    \label{fig:statsVort}
\end{figure}

To validate this interpretation, near-kernel turbulent statistics are used. In the following, the instantaneous vorticity magnitude $\omega$ and strain rate tensor magnitude $S$ averaged over the igniting and failing realizations are computed. Figure~\ref{fig:statsVort} shows $\widehat{\omega_{\mathcal{S}}}(t)$ and $\widehat{\omega_{\mathcal{F}}}(t)$ -- the near-kernel vorticity magnitude average over igniting and failing cases --  for both fuels as a function of time. At $0.4$ ms (when the kernel reaches the mixing layer), higher values of vorticity appear to promote ignition. At that time, the main vortical structure is the vortex ring generated by the kernel, which entrains fuel. It can be concluded that fuel entrainment is an important driver for the ignition of Jet-A. This contrasts with C1, where higher high values of vorticity do not promote ignition. At later times, large values of vorticity are detrimental to ignition for both fuels.

\subsubsection{Cause of ignition failure for different fuels}
\begin{figure}
    \centering
    \includegraphics[width=0.6\textwidth]{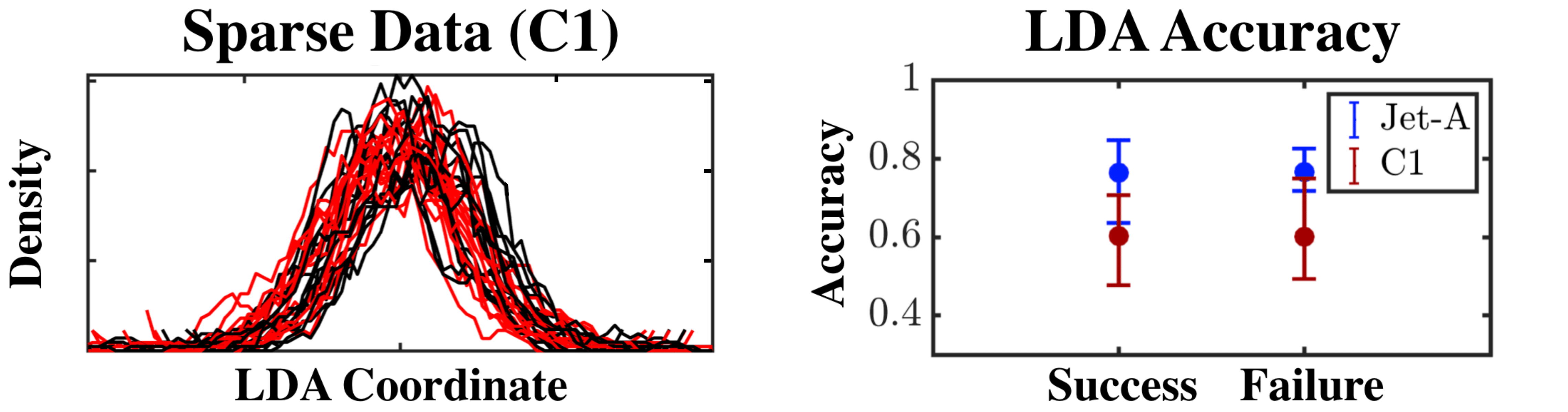}
    \caption{Left: LDA densities of sparse dataset using initial mixture fraction fields for C1 fuel for ignition success (\mythickline{red}) and failure (\mythickline{black}) (analog of sparse data plot in Fig.~\ref{fig:jeta_Z_sensors}). Right: mean LDA accuracies of the full-domain sparse dataset for Jet-A (blue) and C1 (red) for mixture fraction fields at $t=0$. Error-bars indicate maximum and minimum values from the various runs.}
    \label{fig:c1_and_acc}
\end{figure}

% Put figure for classification accuracies --- 
The sensor procedure for the initial mixture fraction field was also carried out for the C1 fuel. The LDA density from the full-domain case, again for 50 realizations of the algorithm, as well as associated classification accuracies for success and failure (along with the accuracy for the full domain case for Jet-A shown in Fig.~\ref{fig:jeta_Z_sensors}) are shown in Fig.~\ref{fig:c1_and_acc}. As opposed to Jet-A, the sparse LDA is significantly less accurate (right plot in Fig.~\ref{fig:c1_and_acc}). The minimum of C1 sparse LDA accuracy for both success and failure reaches the 50\% mark, which is very close to the designed ignition probability of the dataset. Thus, the C1 sensors for initial mixture fraction do not provide any real level of discrimination power. This reveals a key difference in the flow structures that drive ignition between the two fuels, which is interpreted below.

The sparse classification is based on a representation of the LDA vector with a sparse vector. Since representing small scales requires more grid points than large scales, it can be expected that the sparse vector will perform better when representing processes that are dominated by the larger structures in the flow. If the scales that drive the ignition outcome are too small to be efficiently resolved by the sparse vector, then the classification is expected to be ineffective. This motivates the hypothesis that C1 ignition is more influenced by small scales, while Jet-A is influenced by larger structures. 

\begin{figure}
    \centering
    % High res
    %\includegraphics[width=0.6\textwidth,trim={0cm 0cm 2cm 0cm},clip]{strainRateFilt.eps}
    %Low res
    \includegraphics[width=0.6\textwidth,trim={0cm 0cm 2cm 0cm},clip]{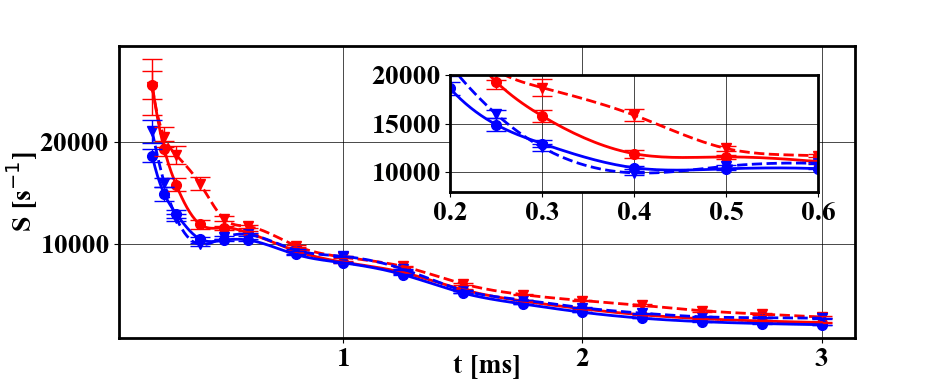}
    \caption{Time evolution of the mean strain rate tensor magnitude at the isosurface $C=0.02$ for Jet-A (blue) and C1 (red) and averaged over the igniting cases  (\mythickErrorcircle{black}{black}) and the non-igniting cases (\mythickErrorVtriangle{black}{black}). Interpolated lines between data points for igniting (\mythickline{black}) and failing cases (\mythickdashedline{black}) are shown to ease visualization. Inset zooms on region of interest.}
    \label{fig:statsStrain}
\end{figure}

To verify this hypothesis, near-kernel strain-rate statistics are examined. This time, the magnitude of the strain rate tensor $S$ is averaged over the flame surface and plotted for all of the igniting and failing cases for both fuels in Fig.~\ref{fig:statsStrain}. Most of the differences between fuels can be observed at $0.4$ ms, which corresponds to the early stages of ignition. At this time, for C1 fuel, the strain rate is about 30\% higher for failing cases than for igniting cases. A similar discrepancy in strain rate for igniting and failing cases for Jet-A is not observed. Vorticity is indicative of an organized turbulence structure while strain rate can be associated with any type of velocity gradient. This supports the fact that an organized structure such as recirculation may be more influential in the case of Jet-A as compared to C1.

%To further understand the exact role of the turbulent scales in inducing ignition failure for C1, another discriminant analysis was performed on the total enthalpy field at several different time instances. Since the resulting sparse LDA accuracy for C1 total enthalpy was found to be low (not shown here), it is concluded that the large-scale features of total enthalpy are similar between failed and successful ignition. It is therefore speculated that the turbulent scales affect C1 ignition not by dissipating heat, but rather by stretching the kernel. 

%% Sensing info for this section:: classification accuracy vs time for C1 for total enthalpy
%\begin{figure}
%    \centering
%    \includegraphics[width=\columnwidth]{c1_htot_acc.eps}
%    \caption{LDA accuracy versus time corresponding to the sparse dataset obtained from total enthalpy fields for C1 fuel. The solid line is the average of 15 runs of the algorithm; upper and lower bounds are maximum and minimum values at each time instant respectively. \alert{MH: if we are low on space, maybe we could just state this.}}
%    \label{fig:c1_htot_sensor}
%\end{figure}

%We should show the layer average just to explain why the downstream matters

\section{Ignition and misfire modes}
\label{sec:ignitionModes}

In the previous section, the types of turbulent structures that affect the ignition outcome (ignition or misfire) were identified in the case of Jet-A and C1 fuels. However, the realization-to-realization variability is not only limited to changes in ignition outcome, but to variation in the process by which the ignition outcome occurs. In other terms, the combustor can experience different ignition and failure modes which can lead to different types of behavior for the combustor. Therefore, knowing what ignition and failure modes are possible in a given combustor is essential for design purposes. In this section, it is shown how the data generated can be used to identify the different ignition and failure modes through a clustering technique (details are provided in Sec.~\ref{sec:postProcModes}). The results are presented and discussed in Sec.~\ref{sec:resultModes}.

\subsection{Post-processing methodology}
\label{sec:postProcModes}

% SB: removed this sentence: Data classification has received much attention in the data science community \cite{aggarwal2014data}. 
To identify ignition and misfire modes, a data-clustering strategy is used. In this section, the data-clustering technique is briefly described. Then it is explained how the clustering is used to identify ignition and misfire modes.

\subsubsection{K-means clustering}

The clustering process is done in an unsupervised manner using K-means clustering \cite{arthur_kmeans,kmeans_review,lloyd1982least}, which has been successfully applied in a variety of fluid applications \cite{barwey2019experimental,barwey2020data,kaiser_crom,dare2019cluster}. The K-means algorithm takes as an input a number of clusters and subsequently groups the realizations (or snapshots) into these many clusters. The groups are created such that the variances of the snapshots within a cluster are minimized. More formally, starting from $N$ data points $\mathcal{X} = \{x_1,...,x_N\}$, one would like to define $N_k$ clusters $\mathcal{C} = \{C_1,...,C_{N_k}\}$. Each cluster $C_k$ is an ensemble of data points and can be characterized by its barycenter, or centroid $c_k = \sum_{x \in C_k} x$. The K-means algorithm solves the following minimization problem 
\begin{equation}
    \argminA_{\mathcal{C}} \sum_{C_k} \sum_{x \in C_{k}} d (x, c_k)^2. \\
\end{equation}
The centroid can then be considered as a representative of its cluster and can be used for interpretation.

\subsubsection{The number of clusters and the norm}

The K-means clustering relies on two ``hyperparameters" that need to be decided by the user: the number of clusters and the type of distance chosen.

There exist indicators that evaluate how suited a candidate number of clusters is to a particular dataset. However, these indicators should not be considered as a systematic way of choosing the number of clusters \cite{barwey2020data}, but rather as guidelines that should be refined by the user. Here, two different indicators (silhouette score~\cite{rousseeuw1987silhouettes} and X-means~\cite{pelleg2000x,kass1995reference,novikov2019pyclustering}) were used to determine the optimal cluster number. Although both methods gave a different cluster number, the results presented in Sec.~\ref{sec:resultModes} were not sensitive to this choice. % (the reader is referred to the Supplementary material for the raw results).
In the results shown, the number of clusters is chosen using the silhouette method, since the number of clusters was lower overall than the X-means counterpart, facilitating interpretation (see Appendix~\ref{app:numberCl}). The silhouette and K-means algorithms are imported from the Scikit-learn library \cite{scikit-learn}, and X-means is imported from the \verb|pyclustering| library \cite{novikov2019pyclustering}.

In the following results, the distance chosen for the clustering is the L$_2$ norm computed in physical space. While this choice has been sufficient in other applications \cite{barwey2019experimental,barwey2020data,kaiser_crom}, it is not necessarily ideal. Other distance measures based on image recognition concepts were investigated \cite{zhang2018unreasonable}, though these led to the same conclusions as the ones presented in Sec.~\ref{sec:resultModes}.

\subsubsection{Interpretation of clusters}

During each ignition or misfire realization, the 3D data is printed at 21 time instances. Then, for each fuel, for each type of event (ignition or misfire), and for each one of the 21 time instances, an optimal number of clusters is computed using the silhouette method. Note that the number of clusters need not be constant across time instances (see Appendix~\ref{app:numberCl}).

For each fuel, for each type of event, and for each one of the 21 time instances, the data is clustered. The result of the clustering is visualized by using the centroid as a representation of each cluster. Since the data clustered is three-dimensional, the centroid is also three-dimensional. Between each time instance, a snapshot jumps from one cluster to another. By recording how often a realization transitioned between clusters, one can build a probability map of the ignition sequence. For example, if 90\% of the realizations that belong to cluster A after 1~ms, belong to cluster B after 2~ms, one can infer that the path from cluster A to B is dominant. This idea is similar to the one developed in Ref.~\cite{kaiser_crom}, except that the goal is not to build a reduced-order model of an ergodic system, but to summarize the evolution of many realizations. The full ignition sequence can be visualized as a forward-in-time network, where vertices are the centroids and edges are the transition probability. The networks allow to quickly visualize different ignition and failure pathways. For clarity, the networks are not reproduced inside the manuscript.%but are provided in the Supplementary material. 

It is emphasized that the number of clusters should be distinguished from the number of ignition modes: a mode refers to a unique physical mechanism that distinguishes it from other modes. In practice, the same mode can appear in several clusters in a more or less pronounced form. Identifying modes from clusters depends on the type of conclusions needed, and requires expert knowledge. The results shown in Sec.~\ref{sec:resultModes} are the result of this analysis. 

\subsection{Results}
\label{sec:resultModes}

\subsubsection{Ignition Modes}
\label{sec:ignModes}

\begin{figure}[h!]
    \centering
    % High res
    %\includegraphics[width=0.49\textwidth,trim={0cm 0cm 0cm 0cm},clip]{C1ProgIgn.eps}
    %Low res
    \includegraphics[width=0.49\textwidth,trim={0cm 0cm 0cm 0cm},clip]{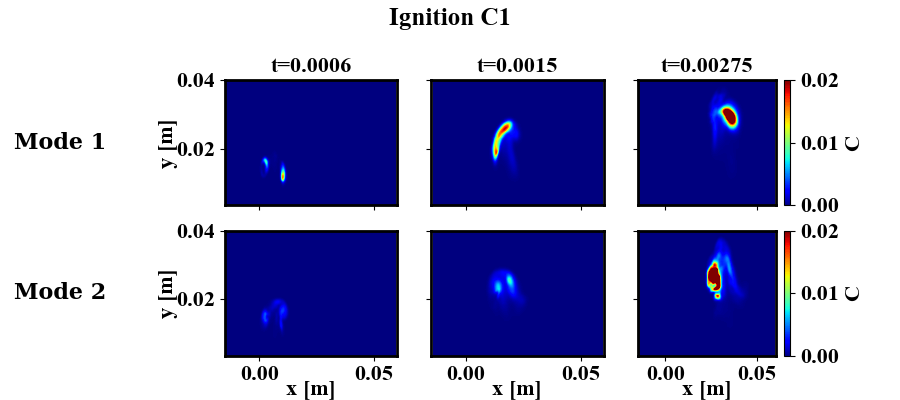}
    %\includegraphics[width=0.49\textwidth,trim={0cm 0cm 0cm 0cm},clip]{C1ProgFail.eps}
    %Low res
    \includegraphics[width=0.49\textwidth,trim={0cm 0cm 0cm 0cm},clip]{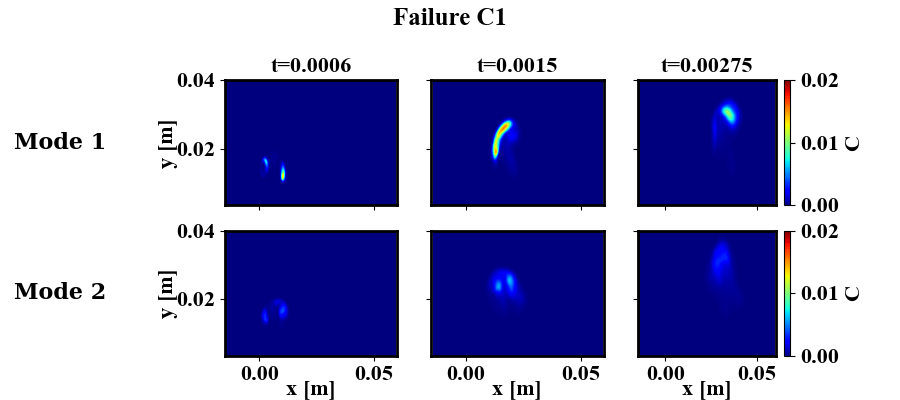}
    \caption{Mid-spanwise planes of progress variable centroids that correspond to Mode 1 (top) and Mode 2 (bottom) for ignition (left) and failure (right) of C1, at $t=0.0006$s, $t=0.0015$s and $t=0.00275$s.}
    \label{fig:C1Ign}
\end{figure}

First, all the progress variable snapshots that correspond to ignition realizations are clustered. Two main ignition paths are found for C1 and are shown in Fig.~\ref{fig:C1Ign} as a sequence of centroids at select time instances. While only the mid-spanwise plane of the centroids are shown here, the clustering is done with the three-dimensional data. The probabilities mentioned here are based on the size of the cluster at the last time instance. In the case of C1, Mode 1 is the most common ignition mode and is found to occur about 95\% of the time (202 snapshots out of 212).  The ignition mode is similar to previous descriptions \cite{jaravel2019numerical}: ignition starts in the counter-rotating vortices of the emerging kernel. Then, the windward part of the vortex ignites while a continuous roll-up of on the leeward vortex mixes the surrounding fuel with the hot kernel. Eventually, ignition develops in the leeward vortex. 

In turn, Mode 2 is found to occur only 5\% of the time (10 snapshots out of 212) and exhibits unique features. As the kernel emerges, counter-rotating vortices are heavily tilted with the windward vortex being located at a lower y-location than the leeward vortex. This behavior leads to a breakdown of the kernel which ignites on average on the windward side of the kernel.

\begin{figure}[h!]
    \centering
    % High res
    %\includegraphics[width=0.49\textwidth,trim={0cm 0cm 0cm 0cm},clip]{JetAProgIgn.eps}
    %Low res
    \includegraphics[width=0.49\textwidth,trim={0cm 0cm 0cm 0cm},clip]{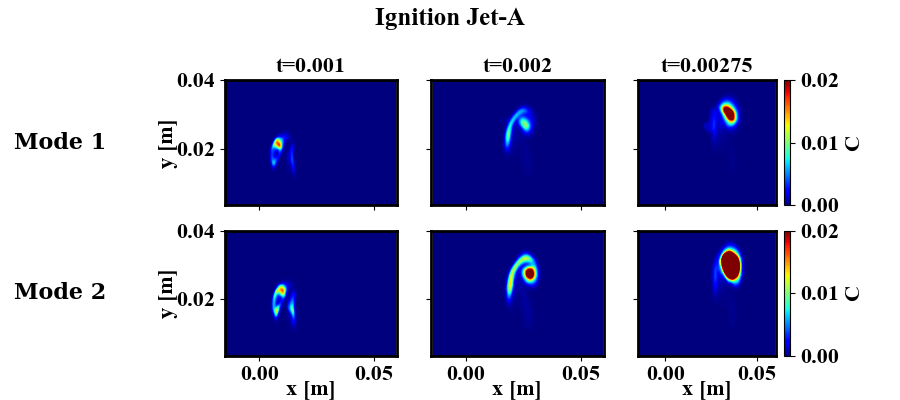}
     % High res
    %\includegraphics[width=0.49\textwidth,trim={0cm 0cm 0cm 0cm},clip]{JetAProgFail.eps}
    %Low res
    \includegraphics[width=0.49\textwidth,trim={0cm 0cm 0cm 0cm},clip]{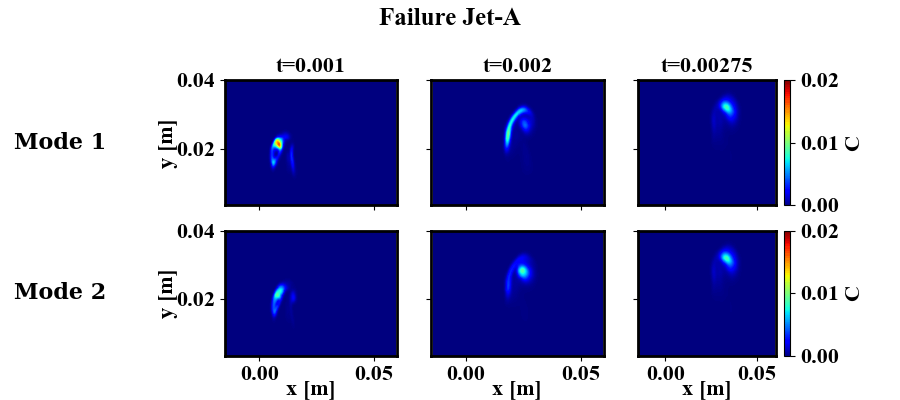}
    \caption{Mid-spanwise planes of progress variable centroids that correspond to Mode 1 (top) and Mode 2 (bottom) for ignition (left) and failure (right) of Jet-A, at $t=0.001$s, $t=0.002$s and $t=0.00275$s.}
    \label{fig:JetAIgn}
\end{figure}

As shown in Fig.~\ref{fig:JetAIgn}, Jet-A also exhibits two main ignition paths, but the differences between these two modes are not as pronounced as for C1. Both modes of ignition of Jet-A resemble Mode 1 of C1 and Mode 2 shows a slightly more pronounced roll-up on the leeward part of the kernel throughout the ignition sequence, which leads to a stronger ignition (higher progress variable). Out of the fuels studied, C1 is the only one that exhibited kernel breakdown which suggests that it is the most affected by small and disorganized turbulence structures. This observation is consistent with the analysis of Sec.~\ref{sec:causalityInference}, where it was found that C1 was influenced by smaller turbulence structures than Jet-A.

\subsubsection{Failure Modes}

By clustering the progress variable field for all the failing ignitions, two main paths are also identified for Jet-A and C1. For C1, the failure modes shown in Fig.~\ref{fig:C1Ign} are similar to the ignition modes. In the case of Mode 1, the progress variable grows first in the windward vortex and then the leeward vortex. However, the progress variable growth is too slow to counteract thermal energy dissipation. In Mode 2, the kernel experience the same tilting as described in Sec.~\ref{sec:ignModes} which later leads to a vortex breakdown. A notable difference is that Mode 2 is observed for 27\% of the snapshots (59 snapshots out of 220), while it was observed for 5\% of the igniting snapshots. When a breakdown occurs (for 69 snapshots), it can be favorable to ignition (10 snapshots) but much more often leads to a misfire (85\% of the time).

In the case of Jet-A, although the paths appear separate, the kernels are in a similar state, as shown in Fig.~\ref{fig:JetAIgn}. Slight differences in the value of the progress variable in the leeward part of the vortex roll-up can be identified. Overall, the values of the progress variable are lower all the way through the misfire sequence compared to the ignition sequence. It can be concluded that the early stages of kernel ignition are instrumental for the ignition outcome. This observation is in line with other work \cite{mastorakos2009ignition,esclapez2015ignition,ahmed2006spark,barre2014flame}.

\subsubsection{Causes and effects of kernel breakdown}
\label{sec:causeEffectBreak}
From the analysis of the previous subsections, it appears that C1 differs from Jet-A because the kernel can experience a breakdown at the later stages of the ignition sequence. Here, the causes and the consequence of the breakdown are further detailed.

\begin{figure}[h!]
    \centering
    % High res
    %\includegraphics[width=0.6\textwidth,trim={0cm 0cm 0cm 0cm},clip]{C1T.eps}
    %Low res
    \includegraphics[width=0.6\textwidth,trim={0cm 0cm 0cm 0cm},clip]{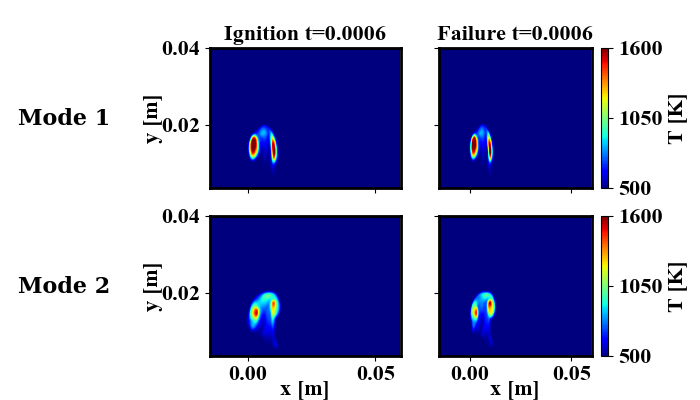}
    \caption{Mid-spanwise planes of temperature centroids that correspond to Mode 1 (top) and Mode 2 (bottom) for ignition (left) and failure (right) of C1, at $t=0.0006$s.}
    \label{fig:C1T}
\end{figure}

First, the breakdown mechanism is investigated by clustering the temperature field following the same procedure as used for the progress variable. Figure~\ref{fig:C1T} shows the mid-spanwise plane of the temperature centroid corresponding to Mode 1 and 2 for the ignition and failure of C1 at 0.6ms. At this stage, the kernel has started tilting and the advancement of the combustion is visibly different. In Mode 1, the windward and leeward vortices (the whole vortex ring) are hotter than in Mode 2 which could act as a viscous buffer that would prevent the kernel from being exposed to turbulent fluctuations.   

While the networks cannot determine the root cause of the kernel tilting, they can be useful for identifying in the dominant flow quantities. One can construct a network of the ignition and failure of C1 for other quantities than the progress variable. For each quantity, the time at which the two ignition modes appear can be used to infer what causes the kernel breakdown. %The networks and the time where the split occurs are shown in the Supplementary Material. 
Table~\ref{t:divergenceTime} shows the time at which the two ignition paths appear for each quantity clustered. Here, the clustering is done for vorticity magnitude ($\omega$), progress variable ($C$), and mixture fraction ($Z$). It can be seen that the earliest path divergence can be observed in the vorticity field while the latest divergence can be observed in the mixture fraction field. It can be concluded that the kernel breakdown is really due to a vortical structure that affects the kernel growth rather than a fuel-feeding mechanism. 

\begin{table}[t]
\begin{center}
\setlength{\abovecaptionskip}{0pt}
\setlength{\belowcaptionskip}{5pt}
\caption{Time instances at which the two ignition and failure path can be distinguished in the ignition network constructed by clustering the vorticity magnitude $\omega$, the progress variable $C$ and the mixture fraction $Z$.}
\label{t:divergenceTime}
\begin{tabular}{c c c c}
\hline
           &  $\omega$ &  $C$ & $Z$          \\ \hline
Ignition   &  $0.08$ms &  $0.3$ms & $0.6$ms  \\ 
Failure   &  $0.08$ms &  $0.25$ms & $0.4$ms  \\  \hline
\end{tabular}
\end{center}
\vspace{0mm} 
\end{table}

\begin{figure}[h!]
    \centering
    % High res
    %\includegraphics[width=0.45\textwidth,trim={0cm 0cm 0cm 0cm},clip]{BreakDownmaxT.eps}
    %Low res
    \includegraphics[width=0.45\textwidth,trim={0cm 0cm 0cm 0cm},clip]{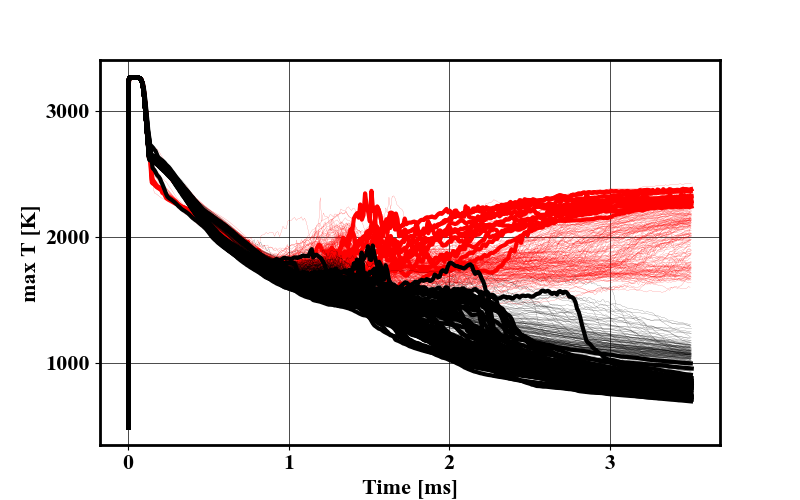}
        % High res
    %\includegraphics[width=0.45\textwidth,trim={0cm 0cm 0cm 0cm},clip]{BreakDownVol.eps}
    %Low res
    \includegraphics[width=0.45\textwidth,trim={0cm 0cm 0cm 0cm},clip]{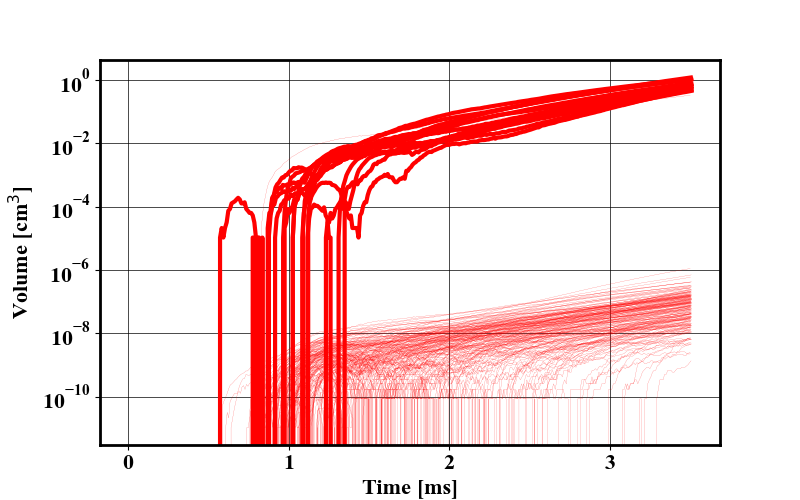}
    \caption{Left: time history of maximal value of temperature over the computational domain for realizations labelled as ignition (\mythickline{red}) and misfires (\mythickline{black}). Right: time history of the volume of product define by the volume of computational cells with $C>0.1$, for igniting cases only. Thin lines denote Mode 1 and thick lines denote Mode 2.}
    \label{fig:breakdown}
\end{figure}

To evaluate the effect of the kernel breakdown, the maximal temperature over the computational domain is tracked for C1 cases ( Fig.~\ref{fig:breakdown}). In cases of ignition, the volume occupied by the products (defined as the volume of computational cells where $C>0.1$), is also tracked (right Fig.~\ref{fig:breakdown}). It can be observed that kernel breakdown (Mode 2) leads to the most extreme ignition and misfire events. If a misfire occurs, it is fastest when the kernel breaks. In cases of a misfire, the reaction rate is too slow to counteract mixing. Therefore, enhanced dissipation leads to faster misfire. Conversely, if ignition starts, it leads to a high maximal temperature and larger fuel consumption. The larger kernel volume is not due to an earlier ignition since products always appear around the same time. Instead, the fuel consumption is more distributed in space. This effect of increasing turbulence was also observed in other studies \cite{yoo2011direct}, where larger integrated heat release was recorded in case of intense turbulence. The larger maximal temperature could be explained by a more intense fuel-kernel mixing which, in this case, overcomes the heat dissipation. In the past, it was found that increasing turbulence could decrease the maximal temperature \cite{chakraborty2007effects}. While this is also true on average here, the finer analysis conducted here reveals that turbulence can lead to high temperatures depending on the flow characteristics.

\section{Summary and conclusions}
\label{sec:summary}

The realization-to-realization ignition variability of Jet-A and C1 observed with multiple large-eddy simulations were analyzed using a range of post-processing methods. The main results are summarized below.
\begin{itemize}
    \item Using discriminant analysis combined with a sparse sensing approach, it was possible to identify that the entrainment of fuel in the kernel is crucial for the ignition of Jet-A.
    \item In the case of C1, the same analysis was not possible because the ignition was found to be dependent on the flame extinction through strain induced by small scale features. This finding highlights how the mechanisms that lead to ignition can be different for different fuels operated in the same geometry. Therefore, the design of aircraft engines regarding altitude relight may need to be revisited when alternative jet fuels are introduced. 
    \item With a data clustering analysis, it was found that two ignition and misfire modes can be observed for C1. One of the modes appears to be the result of an early kernel tilt which later results in kernel breakdown. Two ignition and misfire modes are observed for Jet-A, albeit being similar to each other. The difference lies in the rollup of the leeward part of the vortex ring.
    \item When kernel breakdown occurs for C1, it leads to the most extreme ignition and misfires. During ignition, the breakdown allows a distributed ignition over space and the kernel temperature is larger than when the breakdown does not occur. On average, failure is more likely than ignition when a kernel breakdown occurs.
\end{itemize}

The post-processing methods used here were able to provide a detailed description of realization-to-realization variability for jet fuel relight and can be used in a realistic setting to refine the design of an aircraft engine. When combined with domain knowledge of the combustion process, these data analysis techniques provide extensive insight into the mechanisms that drive flame ignition in such devices. An interesting extension is the application of these techniques to experimental data of gas turbine combustors, which will be pursued in the future.

%{\color{red} We should add some stuff, not sure what. Probably something about the viability of the methods since this is what is announced in the intro.}

\section{Acknowledgements}
This study was supported through a grant from NASA (NNX16AP90A) with Dr. Jeff Moder as the program monitor. The authors also acknowledge the generous allocation of computing time on NASA HECC systems. 

\addcontentsline{toc}{section}{Appendices}
\section*{Appendix}
\renewcommand{\thesubsection}{\Alph{subsection}}

\subsection{Turbulent properties and justification of the realization-to-realization approach}
\label{app:indivReal}

The turbulent properties of the cold flow (initial conditions) are provided for C1 and Jet-A in Tab.~\ref{t:turbProp}. The values are averaged over space and over the ensemble of realizations. The turbulent dissipation rate $\varepsilon = 2 \nu \langle s_{ij} s_{ij} \rangle$, where $s_{ij} = \frac{\partial \overline{U_i}}{\partial x_j}$, and $\overline{U_i}$ is the $i$ component of the filtered velocity, and $\nu$ is the kinematic viscosity. The Kolmogorov timescale is defined as $\tau_{\eta} = \sqrt{\frac{\nu}{\varepsilon}}$, the integral timescale as $\tau_{int} = \frac{k}{\varepsilon}$, and the Taylor microscale Reynolds number is approximated as $Re_{\lambda} = \sqrt{\frac{20}{3} \frac{k^2}{\varepsilon \nu}}$.

\begin{table}[h!]
\begin{center}
\setlength{\abovecaptionskip}{0pt}
\setlength{\belowcaptionskip}{5pt}
\caption{Turbulent properties averaged over space and over the initial conditions. Error bounds indicate root mean square fluctuations over the realizations.}
\label{t:turbProp}
\begin{tabular}{c c c c c c}
\hline
        &  $k$[m$^2$.s$^{-2}$] &  $\varepsilon$ [m$^3$.s$^{-2}$]& $\tau_{\eta}$ [ms]& $\tau_{int}$ [ms]& $Re_{\lambda}$ \\ \hline
C1      &  0.448 $\pm$ 0.032 & 46.38 $\pm$ 3.19   & 2.31 $\pm$ 0.096 & 56.5 $\pm$ 4.9 & 52 $\pm$ 3.7 \\ 
Jet-A   &  0.446 $\pm$ 0.032 & 46.5 $\pm$ 3.1 & 2.33 $\pm$ 0.096 & 55 $\pm$ 4.9 & 50.6 $\pm$ 3.6\\  \hline
\end{tabular}
\end{center}
\vspace{0mm} 
\end{table}

Because of the chaoticity of turbulence, LES cannot, in general, be used to investigate individual realizations \cite{langford-moser}. Over time, any small scale errors including subfilter scale approximations amplify until they dominate the flow field \cite{senoner2008growth,metais-lesieur-predict}. The growth of subfilter scale errors during an ignition realization is analyzed below.

First, errors grow in magnitude at a certain rate (the Lyapunov exponent) \cite{mohan2017scaling,hassanaly2019lyapunov} which should be compared to the simulation time of transient events. An estimate of the rate of growth of perturbations can be obtained from Ref.~\cite{mohan2017scaling}, where a relationship between the perturbation growth rate and $Re_{\lambda}$ is provided. Subfilter scale errors would grow at the exponential rate $e^{42 t}$ for both Jet-A and C1, i.e. would have grown by only $15\%$ after $3.5$ ms. By the end of the simulation, subfilter error magnitudes have marginally grown, which justifies, in part, the realization-to-realization approach.

Second, subfilter errors tend to backscatter over time, and it was found that the characteristic error wavenumber decreases at a rate defined by $\tau_{int}$ \cite{metais-lesieur-predict,leith1972predictability}. For both Jet-A and C1, $\tau_{int}$ is larger than the simulation time. Therefore, it can be expected that subfilter error will only marginally backscatter during the simulations. However small and localized at small scales, if the error is localized at scales that matter for the ignition process, predictions based on initial conditions cannot be achieved. For Jet-A, it was found in Sec.~\ref{sec:causeResults} that large vortical structures affected the ignition process. These large structures are not affected by the backscatter of subfilter scales errors, which explains why it was possible to identify initial conditions that lead to ignition or failure. In the case of C1, it was concluded that small scales affected the ignition process. Since the classification of ignition success and failure based on initial conditions was not possible, and in light of the analysis above, it can be concluded that scales of the size, or smaller than the LES filter size matter, at least initially. At later stages, since the ignition modes are clearly separated (Sec.~\ref{sec:ignModes}) all the way to the end of the simulations, the analysis is unlikely to be influenced by the accumulation of errors over time.

\subsection{Number of clusters}
\label{app:numberCl}

In order to characterize the ignition success and failure mechanism of both fuels, we use the K-means clustering. At each one of the 21 time instance samples during the simulations, clustering is done. Therefore the number of clusters can vary over time.

For the initialization of the clusters, the K-means++ version of the K-means algorithm \cite{arthur_kmeans} is used. The algorithm picks random snapshots as initial centroids of the clusters. This random initialization creates a variance in the silhouette score obtained with a fixed number of clusters. Therefore, the number of clusters is determined by running the silhouette score 21 times for cluster numbers between 2 and 10. The number of clusters selected is the one that gives, on average, the largest silhouette score. This process is repeated for each one of the 21 time instances saved between $t=0$~ms and $t=3.5$~ms, and for each fuel. Therefore, the optimal number of clusters varies over time. 

As mentioned in Sec.~\ref{sec:postProcModes}, a second method is used to determine the optimal number of clusters: the X-means method \cite{pelleg2000x}. This algorithm starts with a certain number of clusters (the parent clusters) and recursively refines them in subgroups (the child cluster). The refined clustering of the parent cluster is then evaluated using a certain criterion \cite{kass1995reference}. If the refined clustering of the parent cluster is deemed more appropriate by the criterion, it is adopted. The clustering initialization is also subject to randomness. Therefore, the optimal number of clusters is also computed 21 times. For each time instance, each field, and each fuel, the average optimal number of clusters is recorded. Note that the clustering is always done with K-means, and X-means is only used to decide on the number of clusters.

\begin{figure}[h!]
    \centering
    % High res
    %\includegraphics[width=0.9\textwidth,trim={0cm 0cm 0cm 0cm},clip]{Ncl_Silhouette.eps}
    %Low res
    \includegraphics[width=0.9\textwidth,trim={0cm 0cm 0cm 0cm},clip]{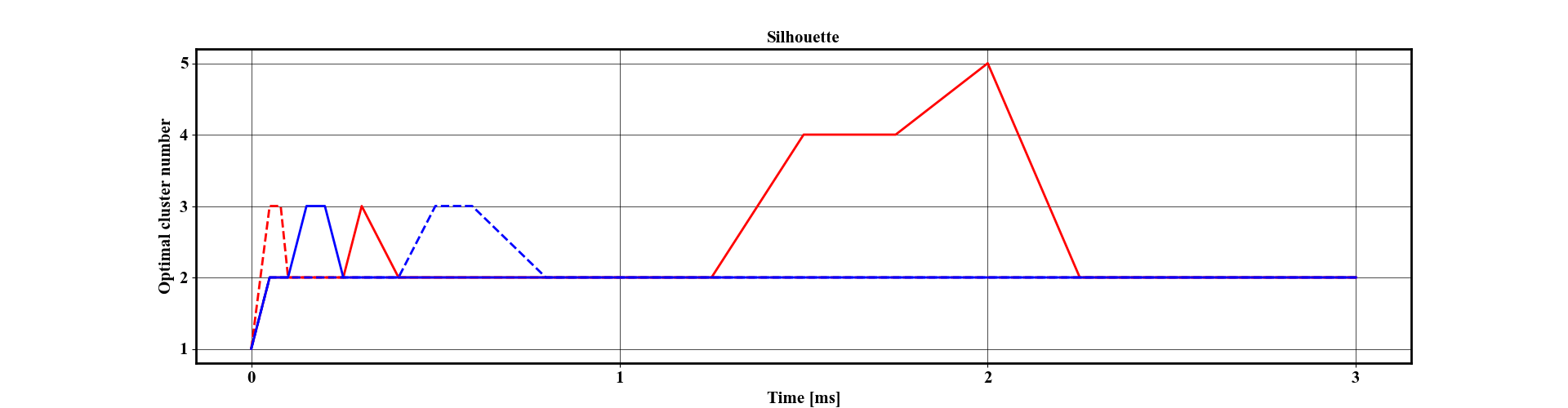}
        % High res
    %\includegraphics[width=0.9\textwidth,trim={0cm 0cm 0cm 0cm},clip]{Ncl_XMeans.eps}
    %Low res
    \includegraphics[width=0.9\textwidth,trim={0cm 0cm 0cm 0cm},clip]{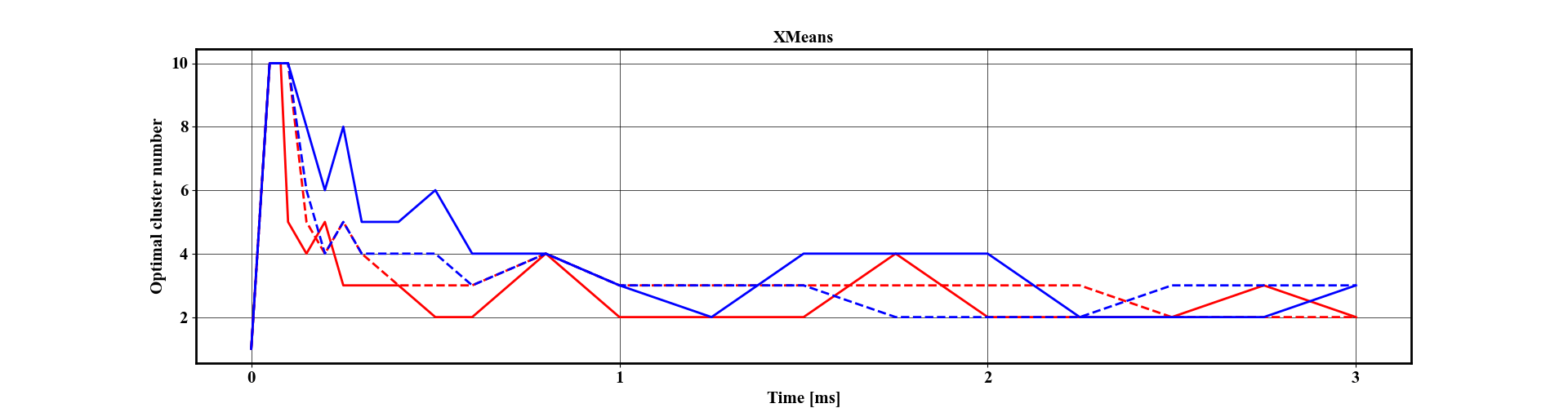}
    \caption{Optimal number of clusters obtained at each time instance for Jet-A (blue) and C1 (red) for igniting (\mythickline{black}) and failing cases (\mythickdashedline{black}), using silhouette score (top), and X-means (bottom).}
    \label{fig:optClust}
\end{figure}

The optimal number of clusters obtained for progress variable over time, for both fuels, are plotted in Fig.~\ref{fig:optClust}. It can be seen that both methods do not agree on the optimal number of clusters but highlight similar trends. At early times, both methods require a larger amount of clusters than at later times. Then, two or three clusters are deemed necessary. The clustering was done with X-means and the silhouette method to decide which method was the most appropriate. The centroid obtained with the X-means did not show different physical processes, which explains why the clustering obtained with the silhouette method was used.%The only way to decide what clustering is appropriate is to plot the resulting networks with both methods. %The results are shown in the Supplementary material. It can be seen that the X-means clustering does not provide more information about the ignition and failure modes, which explains why the clustering obtained with the silhouette score was used.}

\subsection{LDA and Sensor Example}
\label{app:toy_lda}
\begin{figure}[h!]
    \centering
    \includegraphics[width=0.9\columnwidth]{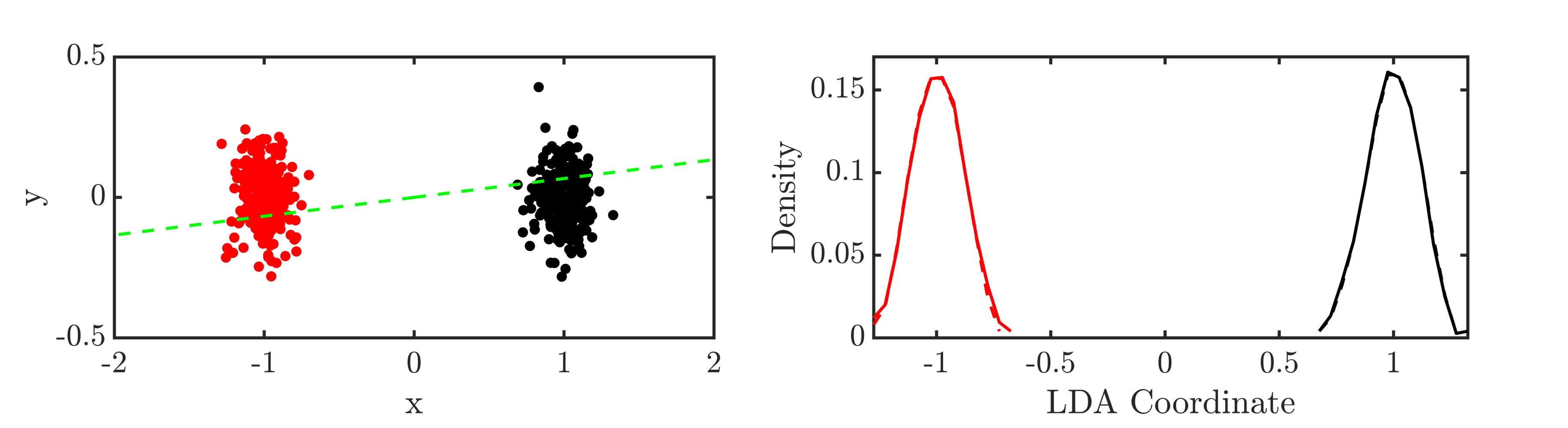}
    \includegraphics[width=0.7\columnwidth]{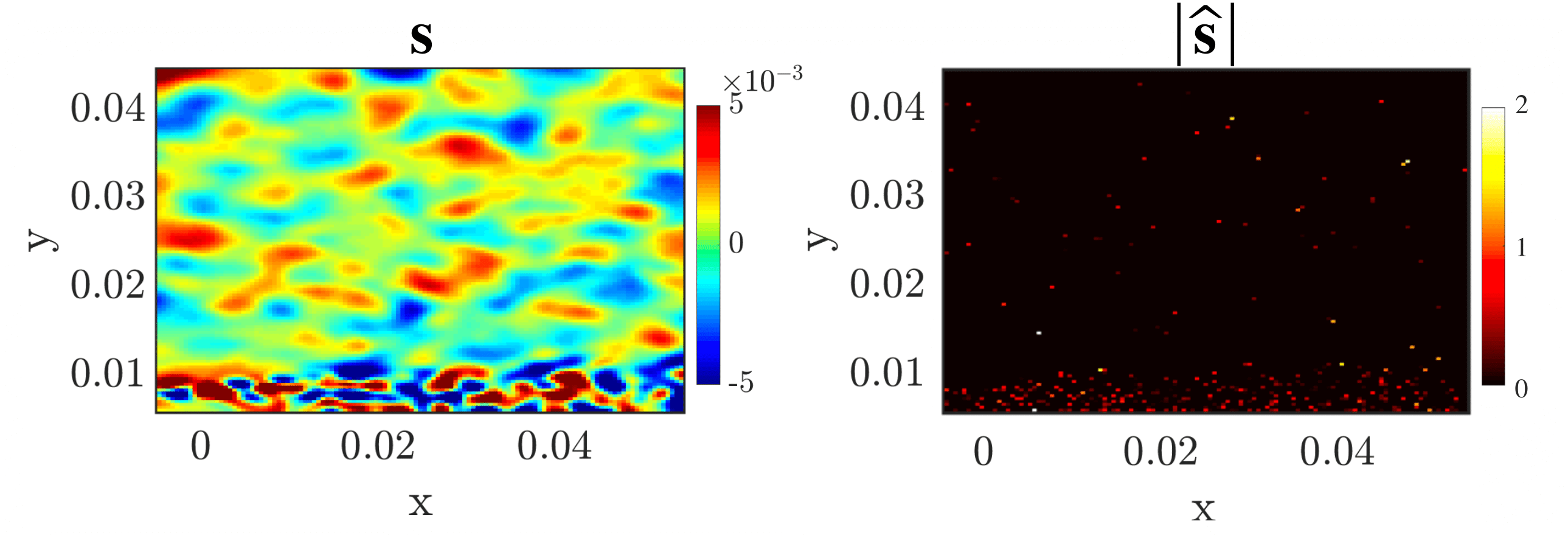}
    \caption{Top left: toy dataset in 2D for success (black) and failure (red), with the LDA vector overlayed (dash-green). Top right: LDA densities for each class for the original data (solid lines) and sparse data (dashed lines). Bottom left: LDA vector in physical space, ${\bf s} = \phi {\bf w}$, from the Jet-A mixture fraction dataset. Bottom right: corresponding sparse sensors ${\bf \hat{s}}$ obtained from the optimization -- the nonzero values are the sensor locations. Both images are X-Y planes obtained at the Z=0 section. As per Eq.~\ref{eq:optim}, both ${\bf s}$ and ${\bf \hat{s}}$ produce the same vector $\bf w$ in POD space.}
    \label{fig:blobs}
\end{figure}

A toy dataset is created in a 2D space by sampling two Gaussian distributions with the following means: $[1,0]$ for the "failure" set, and $[-1,0]$ for the ``success" set. The covariance matrices for both are set to the identity matrix scaled by $0.01$. As shown in Fig.~\ref{fig:blobs} (left), once sampled, the resulting dataset by construction represents two distinct "blobs" separated along the x-axis. Through this trivial example, the goal is to elucidate the meaning of the LDA vector and the sensors. 

Since the data here already exists in such a low-dimensional space (2D), the POD pre-processing step (step 2 in Fig.~\ref{fig:lda_schematic}) is neglected (or, alternatively, the original data and the POD coefficients can be considered identical). Upon labeling the datasets in the same manner as with the LES simulations (Fig.~\ref{fig:blobs}), the LDA can be carried out. The LDA vector was found to be ${\bf w} = [0.998, 0.067]^T$, and is also shown in Fig.~\ref{fig:blobs}. As expected, most of the contribution to ${\bf w}$ comes along the x-axis, which is the direction that best discriminates the two classes. Through this simple example, we can see the significance of the LDA vector: it is a direction that encodes the separation between the two classes from the data given. The projection of the data onto the LDA vector is also illustrated in Fig.~\ref{fig:blobs}, which shows the separation in the LDA coordinates for the two classes in the 1D LDA space. 

We now proceed with recovering the sparse sensors. Recall that in Eq.~\ref{eq:optim}, the optimization problem is defined such that the POD transformation of the sparse vector, $\phi^T{\bf \hat{s}}$, must reproduce the LDA vector $\bf w$, since in the actual implementation, the LDA is performed in the POD space. In this toy example, the original data space and POD space are identical, so the optimization of Eq. 6 is not purposeful (${\bf w} = {\bf s}$). 

Instead of running the optimization, we simply take the best sparse approximation to the already computed $\bf w$. In two dimensions, there are only two options: $[1,0]^T$ or $[0,1]^T$, corresponding to a single sensor in the x- or y-directions, respectively. Clearly, the ideal sensor approximation to ${\bf w}=[0.998, 0.067]^T$ is ${\bf \hat{s}}=[1,0]^T$. Note that in this setting, we only have one sensor ($N_S=1$), and this sensor indicates that only the first dimension (x-direction) is significant in determining the class discrimination. The dashed lines in Fig.~\ref{fig:blobs} (top right) show the LDA projection computed with the sparse data (same concept as in Fig.~\ref{fig:jeta_Z_sensors}), where the sparse dataset was obtained by computing the element-wise product of the original dataset and ${\bf \hat{s}}$ -- since ${\bf \hat{s}}=[1,0]^T$, this amounts to considering only the x-direction and neglecting the y-direction from the original Gaussian data, and then projecting this modified ``sparse" dataset onto the LDA vector $\bf w$. The densities almost exactly overlap, indicating the reliability of the sensor approximation. 

The same concept applies in higher dimensions as used in the ignition dataset, where each dimension in the $M$-dimensional phase space represents an (x,y,z) location in physical space. An example of an actual LDA vector obtained from the mixture fraction dataset juxtaposed with the corresponding sensors resulting from one optimization run in Eq.~\ref{eq:optim} is shown in Fig.~\ref{fig:blobs} (bottom). Each of these plots is analogous to the 2D toy problem counterparts. In the end, the sparse sensors can be regarded as a more interpretable proxy for the LDA vector $\bf w$.  

It should be noted that in the bottom plots of Fig.~\ref{fig:jeta_Z_sensors}, the \textit{frequency} at which a sensor occupies a particular location in physical space is plotted. These frequencies are obtained by performing several runs of the optimization from the same dataset, as described in Sec.~\ref{sec:lda}.

\clearpage
%\bibliography{master.bib}
%\bibliographystyle{elsarticle-num}
\end{document}